\documentclass[10pt,a4paper]{article}

\usepackage{amsmath}
\usepackage{graphicx}
\usepackage[usenames,dvipsnames]{color}
\usepackage{authblk}
\usepackage[left=3.5cm,right=3.5cm,top=3cm,bottom=3cm]{geometry}

\newcommand{\vect}[1]{\mbox{\boldmath $#1$}}

\begin{document}

\title{Temporal interactions facilitate 
endemicity in the susceptible-infected-susceptible epidemic model}

\author[1]{Leo Speidel}
\author[2,3]{Konstantin Klemm}
\author[3]{V\' ictor M. Egu\' iluz}
\author[4,\thanks{email: naoki.masuda@bristol.ac.uk}]{Naoki Masuda}

\affil[1]{Doctoral Training Centre in Systems Biology, University of Oxford, Rex Richards Building, South Parks Road, Oxford OX1 3QU, United Kingdom}
\affil[2]{School of Science and Technology, Nazarbayev University, Qabanbay Batyr Ave 53, Astana 010000, Kazakhstan}
\affil[3]{Instituto de F\'isica Interdisciplinar y Sistemas Complejos IFISC (CSIC-UIB), E07122 Palma de Mallorca, Spain}
\affil[4]{Department of Engineering Mathematics, University of Bristol, Merchant Venturers Building, Woodland Road, Clifton, Bristol BS8 1UB, United Kingdom}

\date{}

\maketitle

\begin{abstract}
Data of physical contacts and face-to-face communications suggest temporally varying networks as the media on which infections take place among humans and animals. Epidemic processes on temporal networks are complicated by complexity of both network structure and temporal dimensions. Theoretical approaches are much needed for identifying key factors that affect dynamics of epidemics.
In particular, what factors make some temporal networks stronger media of infection than other temporal networks is under debate. We develop a theory to understand the susceptible-infected-susceptible epidemic model on arbitrary temporal networks, where each contact is used for a finite duration. We show that temporality of networks lessens the epidemic threshold such that infections persist more easily in temporal networks than in their static counterparts. We further show that the Lie commutator bracket of the adjacency matrices at different times is a key determinant of the epidemic threshold in temporal networks. The effect of temporality on the epidemic threshold, which depends on a data set, is approximately predicted by the magnitude of a commutator norm.\\

\end{abstract}

\section{Introduction}

A majority of infectious diseases, ranging from seasonal influenza to Ebola outbreaks and sexually transmitted infections, can be viewed to occur on contact networks of humans and animals, which are composed of individuals and dyadic links between them. Epidemic processes are one of the most widespread applications of network analysis. Structure of contact networks has been shown to affect, for example, the likelihood and speed of an infection penetrating into a significant part of a population, effectiveness of intervention strategies, and identification of super-spreading individuals \cite{Keeling2005JRSocInterface,Barrat2008book,Pastor-Satorras_review2015}.

Accumulating data evince that contact networks underlying epidemic and other processes are often highly dynamic, constituting temporal networks \cite{HolmeSaramaki2012PhysRep,Holme2015EurPhysJB}. For example, links may be only occasionally used for actual physical contacts. Individuals may be socially active in some restricted periods of time. Temporality of networks may alter effects of networks on epidemic processes \cite{Bansal2010JBiolDyn,Masuda2013F1000Prime}. This is a practical enquiry because various instances of epidemics in human and animal populations, and also viral spreading of information in human society, seem to occur on temporally varying networks. However, our understanding of epidemic processes in temporal networks is still limited. Theory based on the branching process enables us to understand long-tail behaviour of the number of newly infected individuals \cite{Vazquez2007,Iribarren2009PhysRevLett,Iribarren2011PhysRevE}. Other theoretical approaches include analysis of epidemic spreading on theoretically tractable generative models of temporal networks \cite{Eames2004,Volz2007ProcRSocB,Volz2009JRSocInterface,Gross2006PRL,Shaw2008PRE,Schwarzkopf2010PRE,Perra2012,Taylor2012,GuoTrajanovski2013PhysRevE,Ferreri2014,Liu2014}. However, these studies assume network models such that they may miss effects of the properties of temporal contact networks that are present in empirical data but not modelled. On empirical temporal networks, the most popular approach has probably been to run numerical simulations of epidemic processes on the empirical networks and their variants (e.g., \cite{Karsai2011PhysRevE,Miritello2011PhysRevE,Rocha2011PlosComputBiol,Masuda2013F1000Prime}).

Valdano and colleagues introduced a temporal-network variant of the individual-based approximation to understand the susceptible-infected-susceptible (SIS) model of epidemic spreading~\cite{Valdano2015PhysRevX,Valdano2015arxiv} (also see \cite{RochaMasuda2015arxiv} for a similar approach to a different disease model). In this approach, the probability that each node is infected is tracked over time using the matrix algebra. They showed how to calculate the epidemic threshold (i.e., strength of infection above which infection can remain prevalent in the population) and the prevalence (i.e., fraction of infected nodes in the stationary state) in terms of the spectral radius of a relevant matrix. Their theory is applicable to arbitrary temporal network data.

Despite these and other developments, we are yet left with fundamental questions regarding mechanisms of infections in given temporal networks. Why is the epidemic threshold large in one temporal network and small in another? How do two time scales, one of network dynamics and the other of the epidemic process, interact?
In the present study, we use the individual-based approximation to reveal factors controlling the epidemic threshold for the SIS model on arbitrary temporal networks. We show theoretical and numerical evidence that the epidemic threshold decreases (i.e., infection is more likely to occur) as the network becomes more temporal in the sense that the network changes more slowly (but not quiescent) relative to the time scale of the epidemic process. We use the 
continuous-time SIS model on networks switching with regular intervals to reach this conclusion. This result is consistent with those derived for particular temporal network models \cite{Schwarzkopf2010PRE,Taylor2012}. Impacts of the temporality of networks on the epidemic threshold vary across networks \cite{Valdano2015PhysRevX,Valdano2015arxiv,RochaMasuda2015arxiv}. We find that non-commutativity of the adjacency matrices at different times, as quantified by the Lie bracket of the adjacency matrices, is a key indicator that influences the epidemic threshold.

\begin{figure}
\centering
\includegraphics[width=0.7\textwidth]{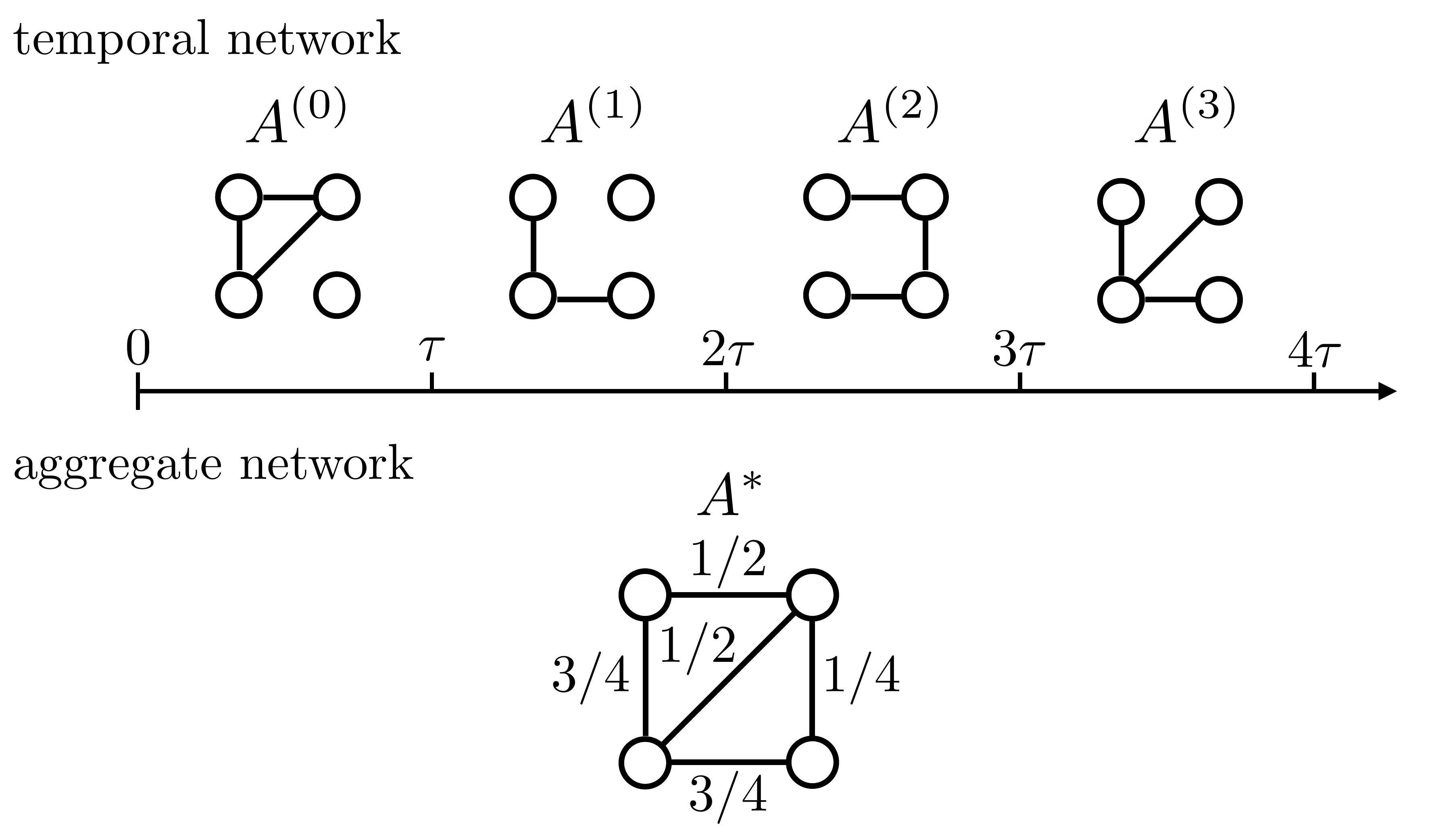}
\caption{Schematic showing a temporal network with $N=4$ nodes and $\ell=4$ snapshots, and the corresponding aggregate network. The link weight in the aggregate adjacency matrix, $A^*$, is equal to the sum of the link weight over the four snapshots divided by four.}
\label{fig:schem}
\end{figure}

\section{Model}

We consider the continuous-time SIS model on undirected temporal networks having $N$ nodes, as schematically shown in Fig.~\ref{fig:schem}. Each node assumes either the susceptible or infected state. An infected node infects each of its susceptible neighbours at rate $\beta$. An infected node transits to return to the susceptible state at rate $\gamma$.
To model exogenous dynamics of networks, we consider an infinite sequence of adjacency matrices $\{A^{(0)}, A^{(1)}, \ldots\}$ and sequentially apply each of them for time $\tau$. In other words, $A^{(\ell^{\prime})}$ is applied between time $\ell^{\prime}\tau$ and $(\ell^{\prime}+1)\tau$. We refer to each network applied for time $\tau$ as the snapshot network, or snapshot in short.

This switching network modelling of temporal networks is common in studying synchronisation processes \cite{Liberzon2003book,Olfatisaber2007IEEE,Masuda2013PhysRevLett}. It is also in accordance with observation of temporal network data at regular time intervals $\tau$. Here we regard $\tau$ as a free parameter. It controls the relative time scale of the epidemic and network dynamics; multiplying $\tau$ by a constant $c (>0)$ is equivalent to not changing $\tau$ and instead changing
$\beta$ and $\gamma$ to $c\beta$ and $c\gamma$, respectively.
In addition, changing ($\tau$, $\beta$, $\gamma$) to ($c\tau$, $\beta/c$, $\gamma/c$) does not change the dynamics. Therefore, we set $\gamma=1$ without loss of generality.

The model simplifies over real epidemic processes. Heterogeneities are considered only through the quenched disorder of the network. A more refined model would account for heterogeneity also in the parameters, drawing $\beta$ and $\gamma$ separately for each node and $\tau$ for each snapshot from probability distributions.  We suppress these refinements for the sake of tractability of the model.

\section{Infections persist more easily in temporal than static networks}

\begin{table}

\centering
\begin{tabular}{lrrrrc}
\hline
Data																											& $N$ & $M$ & $M_{\rm event}$ & $r$ & Aggregating window\\		
\hline
hospital ward~\cite{Vanhems2013}		&		75			&	1,139	&	32,424		&				97			&		1 hour			\\
reality	 mining~\cite{Eagle2009}					&	104			&	3,525	&	781,653	&				54			&		7 days		\\
ht09	~\cite{Isella2011}								      	&		113		&	2,196	&	20,818		&				59			&		1 hour			\\
school2011~\cite{Fournet2014}	    	&		126		&	1,710	&	28,561		&			76				&		1 hour			\\
school2012~\cite{Fournet2014}				&		180		&	2,220	&	45,047		&			203			&	  1 hour			\\
school2014~\cite{Stehle2011,Gemmetto2014}&  242  	&		8,317	&125,773	  &			33				&		1 hour			\\
school2013~\cite{Mastrandrea2015}	&		327		&	5,818	&188,508 	&			101	  	&		1 hour			\\
online message~\cite{Opsahl2009}     & 1,892  & 13,835	& 59,831   &   195   & 1 day\\
hospital~\cite{Gracia-arXiv}								   &	5,607		&	60,177&	936,101 &			105			&		7 days		\\
sexual contact~\cite{Rocha2010}       & 15,810&	38,540& 50,116   & 	 75 	  	& 30 days\\
\hline
\end{tabular}
\caption{Properties of empirical temporal networks. The number of nodes ($N$), that of links ($M$), that of events ($M_{\rm event}$), and that of snapshots ($r$) are shown. The largest connected component of the aggregate undirected network is used for each data set. The length of the aggregating time window used for Figs.~\ref{fig:empirical} and \ref{fig:plot_Cdependency} is also shown.
In the reality mining data set, we have ignored the first 27 weeks, because the time stamps for these entries are false~\cite{Eagle2009}.
\label{tab:data}}
\end{table}

\begin{figure}
\centering
\includegraphics[width=\textwidth]{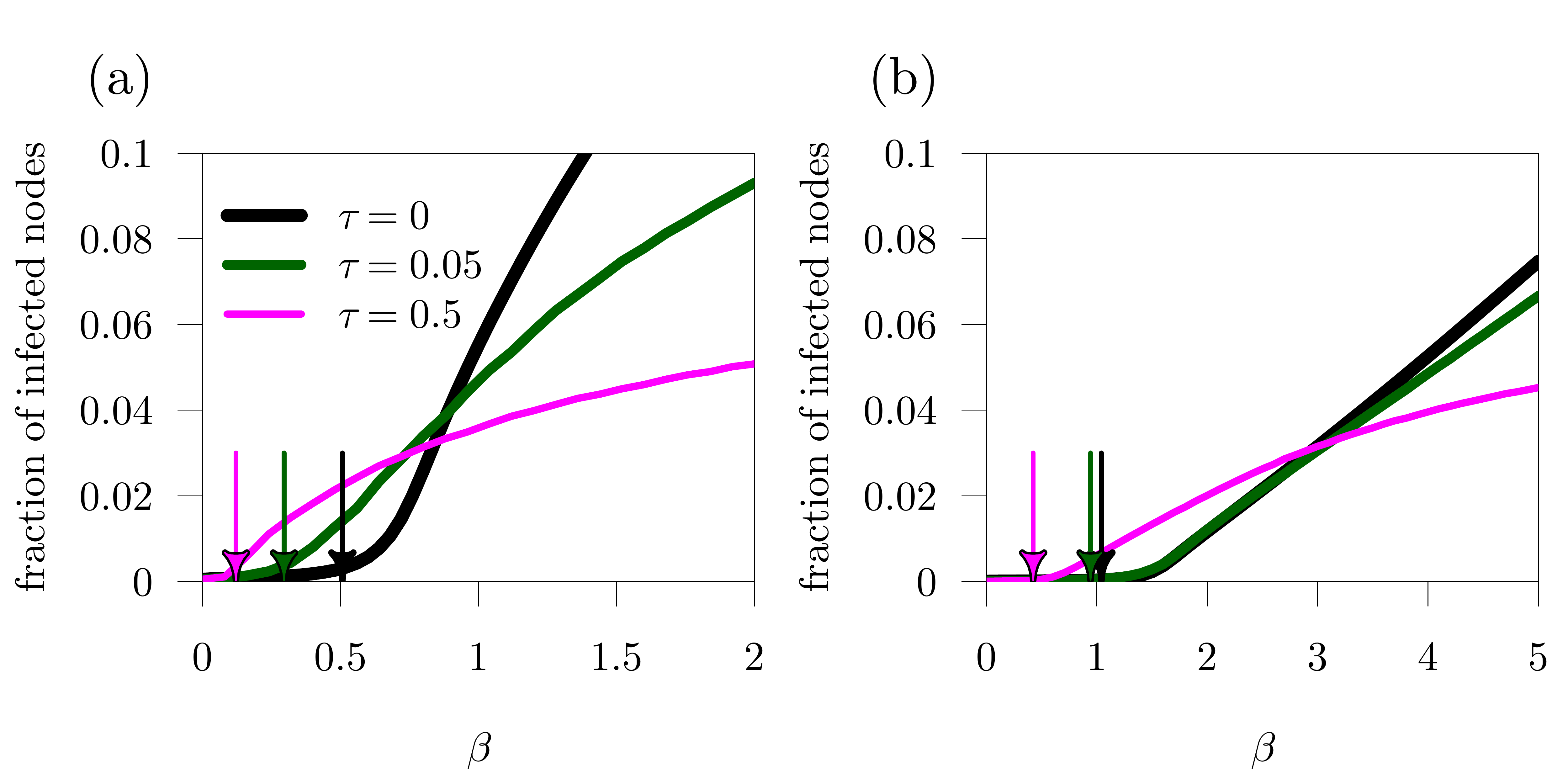}
\caption{Stochastic simulations of the SIS model on the (a) online message data set~\cite{Opsahl2009} and (b) sexual contact data set~\cite{Rocha2010}. See Table~\ref{tab:data} for details on the data sets.
We imposed periodic boundary conditions. The theoretically obtained $\beta_{\rm c}$ is shown by the arrows. We calculated the theoretical $\beta_{\rm c}$ value by a bisection method on the leading eigenvalue of $T(\tau)$. We expanded $T(\tau)$ up to the term of the order of $O(\tau^{10})$ and calculated its leading eigenvalue using the power method.
\label{fig:empirical}}
\end{figure}

We denote the probability that node $i$ ($1\le i\le N$) is in the infected state at time $t$ by $x_i(t)$. Under the assumption that the states of different nodes are independent of each other, the individual-based approximation to 
the SIS dynamics linearised around the disease-free configuration (i.e., $x_i(t)=0$ for all $i$) is given by
\begin{equation}
\dot{\vect{x}}(t) = (\beta A^{(\ell^{\prime})} - I) \vect{x}(t) \quad (\ell^{\prime} \tau \le t < (\ell^{\prime} +1)\tau),
\label{eq:SISonnets_mat}
\end{equation}
where $\vect{x}(t) = (x_1(t),\ldots,x_N(t))^\top$, $\top$ denotes the transposition, and $I$ is the identity matrix.
When the network is static, Eq.~\eqref{eq:SISonnets_mat} is simplified to $\dot{\vect{x}}(t) = (\beta A-I)\vect{x}(t)$,
where $A$ is the adjacency matrix. The epidemic threshold, denoted by $\beta_{\rm c}$, is defined as the value of $\beta$ above which infection can persist in the network. According to the individual-based approximation, $\beta_{\rm c}$ for a static network is given by the value of $\beta$ at which the leading eigenvalue of $\beta A-I$ is equal to zero. The leading eigenvalue of $\beta A -I$ is given by $\beta \alpha_{\max}-1$, where $\alpha_{\max}$ is the leading eigenvalue of $A$. Therefore, we obtain
$\beta_{\rm c} = 1/\alpha_{\max}$ \cite{Wang2003,Castellano2010,Pastor-Satorras_review2015}.

When the network varies in time, Eq.~\eqref{eq:SISonnets_mat} yields
\begin{equation}
\vect{x}(\ell \tau) = T(\tau) \vect{x}(0),
\end{equation}
where
\begin{equation}
T(\tau) = \exp\left[(\beta A^{(\ell-1)}-I) \tau\right] \cdots \exp\left[(\beta A^{(0)}-I) \tau\right].
\label{eq:T(S;tau),general}
\end{equation}
The leading eigenvalue of $T(\tau)$, which depends on sequence $\{A^{(0)}, \ldots, A^{(\ell-1)}\}$, is denoted by $\mu_{\rm max}$. The epidemic threshold, $\beta_{\rm c}$, satisfies $\mu_{\max} = 1$. Valdano and colleagues were the first to derive this result by analysing the SIS model in discrete time \cite{Valdano2015PhysRevX}.

For two temporal networks, we simulated the SIS model dynamics using the quasistationary state (QS) method \cite{Oliveira2005} tailored to the case of temporal networks (Appendix~\ref{sec:QS method}). We aggregated the original data over several time windows to define snapshot networks $A^{(\ell^{\prime})}$; the size of the aggregating window is shown in Table~\ref{tab:data}. The average prevalence of infection for values of $\beta$ and $\tau$ is shown in Fig.~\ref{fig:empirical}. Similarly to Ref.~\cite{Valdano2015PhysRevX}, we assumed a periodic boundary condition such that the first snapshot ensues after the last snapshot. 
The theoretical estimates of $\beta_{\rm c}$ are shown by the arrows in Fig.~\ref{fig:empirical}(a) and (b), which are fairly accurate.

Figure~\ref{fig:empirical} suggests that the epidemic threshold, denoted by $\beta_{\rm c}$, decreases as $\tau$ increases in both networks. To formulate this point, we compare $\beta_{\rm c}$ with the epidemic threshold for the SIS dynamics occurring on the aggregate, static network, denoted by $\beta_{\rm c}^*$. 
We start with defining the aggregate network as the adjacency matrix
$A^*= \sum_{\ell^{\prime}=0}^{\ell-1} A^{(\ell^{\prime})} / \ell$ applied for $0\le t\le \ell \tau$. With this normalisation, the temporal and aggregate networks have the same time average of the weight of each link \cite{Masuda2013PhysRevLett} (Fig.~\ref{fig:schem}). The epidemic threshold for the aggregate network is given by $\beta_{\rm c}^* = 1/\alpha_{\max}^*$, where $\alpha_{\max}^*$ is the largest eigenvalue of $A^*$. 

Any real-valued, continuous spectral function $\phi$ that acts on the spectrum of its matrix argument and only attains finite values satisfies $\phi(e^{M_1}e^{M_2}) \ge \phi(e^{M_1+M_2})$ for arbitrary symmetric matrices $M_1$ and $M_2$~\cite{Cohen1982}. Although a generalisation of this inequality to the case of more than two matrices is false in general, we conjecture that it remains true if the involved matrices have only non-negative entries and $\phi$ is the spectral radius. We have the following evidence. First, to the best of our knowledge, all counterexamples involve matrices with negative entries~\cite{Thompson1965}. Second, the theoretical results for two model temporal networks presented in Sec.~\ref{sec:modelnetwork} are consistent with this inequality. Third, all numerical calculations performed in this paper on real data sets and synthetic networks are consistent with this inequality. By admitting this generalised inequality, and applying it to matrices $\beta \tau A^{(\ell^\prime)}$ ($0 \le \ell^\prime \le \ell - 1$), we obtain
\begin{align}
\mu_{\max} &=
\phi \left( \exp\left[(\beta A^{(\ell-1)}-I)\tau\right] \cdots \exp\left[(\beta A^{(0)}-I)\tau\right] \right) \nonumber\\ 
& = \exp\left(-\ell\tau\right) \phi \left( \exp\left[\beta \tau A^{(\ell-1)} \right] \cdots  \exp\left[\beta \tau A^{(0)} \right]
\right) \nonumber\\ 
& \ge \exp\left(-\ell\tau\right) \phi \left( \exp\left[\beta \tau \sum_{\ell^{\prime}=0}^{\ell-1} A^{(\ell^{\prime})} \right] \right) \nonumber\\ 
& =  \phi \left( \exp\left[\left(\beta A^*- I \right) \ell \tau \right]\right).
\label{eq:mu_max_general}
\end{align}
%
Equation~\eqref{eq:mu_max_general} implies that $\mu_{\max}\ge 1$ at $\beta=\beta_{\rm c}^*$. Therefore, $\beta_{\rm c} \le \beta_{\rm c}^*$.
%

Several remarks are in order. First, Eq.~\eqref{eq:mu_max_general} implicitly assumes the periodic boundary condition for the temporal networks. However, because $\ell$ is arbitrary, Eq.~\eqref{eq:mu_max_general} holds true for arbitrary sequences of networks.
Second, in the limit $\tau \to 0$, Eq.~\eqref{eq:mu_max_general} is satisfied with equality such that $\beta_{\rm c} = \beta_{\rm c}^* = 1/\alpha_{\rm max}^*$. Third, if each snapshot is composed of a single link, numerical results suggest that the epidemic threshold increases as the temporality of the network increases (Appendix~\ref{sec:1 edge}), which is opposite to the current result. In this situation, the probability that the infection is extinguished is not negligible even for a large infection rate. A theory that accounts for stochasticity, which is different from the deterministic individual-based approximation, correctly predicts the direction of the change in the epidemic threshold as $\tau$ increases (Appendix~\ref{sec:1 edge}). It should be noted that a network mainly composed of isolated single links is realistic for sexually transmitted infections through monogamous relationships~\cite{Eames2004}. This situation is out of the scope of the following analysis. Our theory requires that each snapshot has a relatively large connected component such that the SIS dynamics on the snapshot are not significantly influenced by disconnected single links even if they are present. The stochasticity and the absorbing configuration (all nodes susceptible) become relevant not only for disconnected single links (Appendix~\ref{sec:1 edge}) but also snapshots of increasing size as $\tau \to \infty$. If $\tau$ exceeds the typical time of reaching the absorbing configuration on a snapshot, the epidemic ceases due to stochasticity.

\section{Analysis of model networks\label{sec:modelnetwork}}

\subsection{Temporal networks with clique snapshots\label{sub:clique}}

To investigate the distance between the epidemic threshold for aggregate and temporal networks and its dependence on parameters, we start by calculating the epidemic threshold for two temporal network models. In the first model, snapshots consist of a disjoint union of cliques, each with $d_{\rm cl}+1$ nodes, and isolated nodes. 
A clique may be a suitable model for conversation events in a small group~\cite{Miritello2011PhysRevE,Stehle2010,Tantipathananandh2007,Zhao2011}.
The clique size remains the same across different cliques and snapshots. The number of nodes stays constant across snapshots, but the number of cliques may depend on a snapshot.

\begin{figure}
\centering
\includegraphics[width=\textwidth]{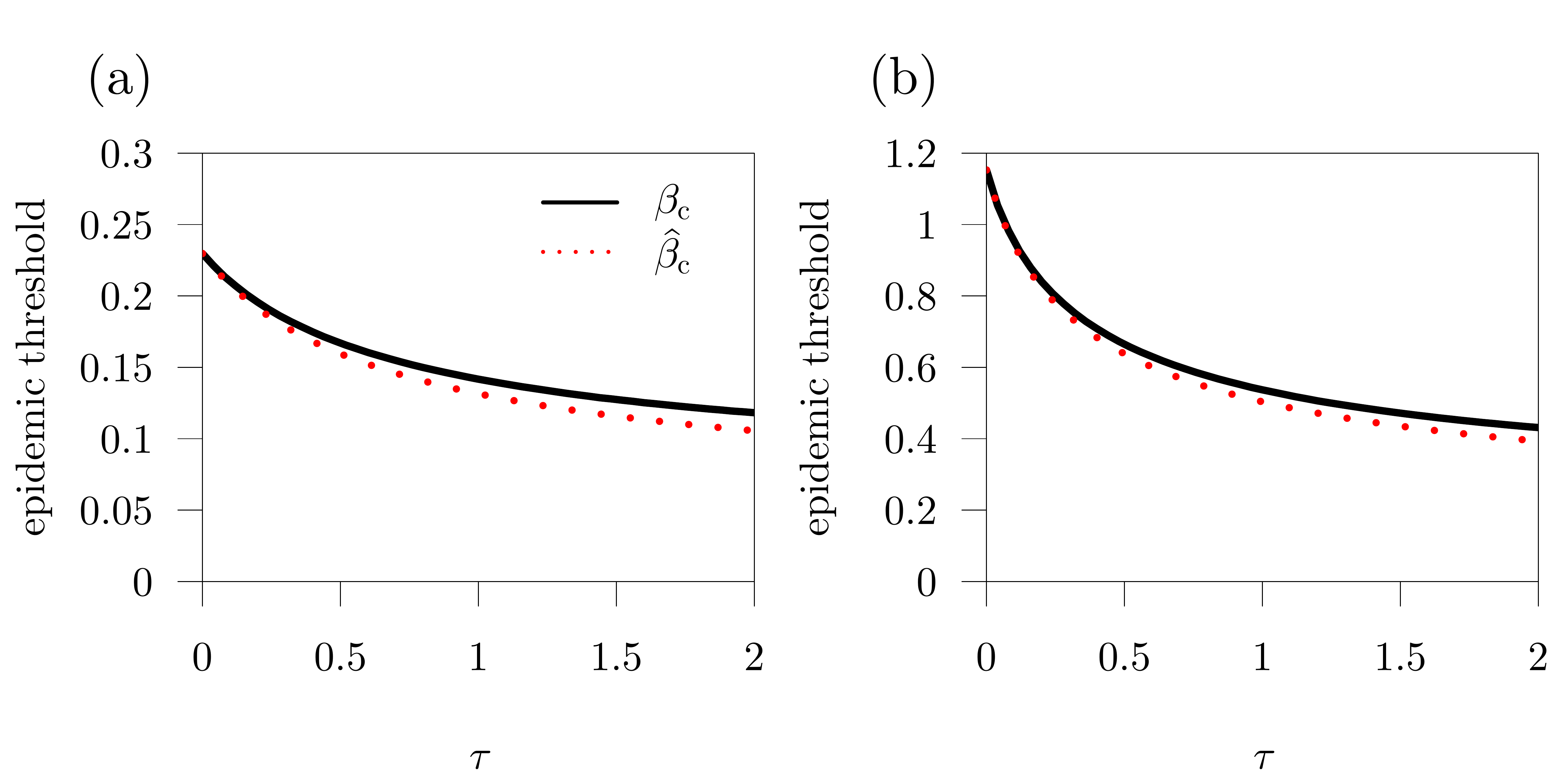}
\caption{Comparison between the epidemic threshold $\beta_{\rm c}$ (solid line) and its lower bound $\hat{\beta}_{\rm c}$ (dotted line). The Lyapunov exponent, which was used for calculating $\hat{\beta}_{\rm c}$, was computed with a Monte Carlo algorithm~\cite{CrisantiBook1993}. (a) Network with random clique snapshots. (b) Activity driven model. We set $N = 100$, $d_{\rm cl} = d_{\rm hub} = 10$, and $r=100$. The value of $a_i$ ($\varepsilon \le a_i\le 1$) is distributed according to a power law with exponent equal to $3$. We adjust $\varepsilon$ such that the mean of $a_i$ equals $0.04$.
\label{fig:lyapunov}
}
\end{figure}

We assume that snapshots are randomly and independently drawn from a set of possible snapshots with equal probability, which we call the random sampling with replacement.
Whether the expected prevalence is positive or not can be determined by $\lambda = \lim_{\ell \to \infty} (\ell \tau)^{-1} \ln \mu_{\rm max}$, which is equal to the maximum Lyapunov exponent associated with the switching linear dynamics induced by operator $T(\tau)$
(Eq.~\eqref{eq:T(S;tau),general}) as $\ell\to\infty$~\cite{Furstenberg1960,CrisantiBook1993,Protasov2013}. The epidemic threshold corresponds to $\lambda = 0$. It is algorithmically undecidable to determine whether $\lambda < 0$ or not~\cite{Tsitsiklis1997}, hindering us from deriving the exact value of $\beta_c$. Therefore, we evaluate the expected state vector, $\mathbf{E}\left[\vect x(t)\right]$, where
$\mathbf{E}$ is the expectation. The expected state vector evolves according to
\begin{equation}
\mathbf{E}\left[\vect x((\ell^{\prime}+1)\tau)\right] = \hat{T}(\tau) \mathbf{E}\left[\vect x(\ell^{\prime}\tau)\right],
\end{equation}
where $\ell^{\prime}=0, 1, \ldots$, and 
\begin{equation}
\hat{T}(\tau) = \frac{1}{r} \sum_{A} \exp\left[(\beta A - I) \tau\right].
\label{eq:effective_matrix,temporal}
\end{equation}
Here, the summation runs over all possible snapshots, and $r$ represents the number of possible snapshots.
We denote the leading eigenvalue of $\hat{T}(\tau)$ by $\hat{\mu}_{\rm max}$; $(\ln \hat{\mu}_{\rm max})/\tau$ approximates $\lambda$. It holds true that $(\ln \hat{\mu}_{\rm max})/\tau \ge \lambda$ (Appendix~\ref{sec:Lyapunov proof}). In addition,
we numerically verified $\beta_{\rm c} \approx \hat{\beta}_{\rm c}$ for some networks and a range of parameter values, where the estimate of the epidemic threshold $\hat{\beta}_{\rm c}$ is obtained from $\hat{\mu}_{\max}=1$ (Fig.~\ref{fig:lyapunov}).

As shown in Appendix~\ref{sec:epidemic threshold clique}, we obtain
\begin{equation}
\hat{\beta}_{\rm c} \approx \frac{1}{\tau d_{\rm cl} } \ln \left[ 1 + \frac{d_{\rm cl}}{ \alpha_{\max}^* }  (e^\tau -1) \right].
\label{eq:hatbeta_c,clique}
\end{equation}
Equation~\eqref{eq:hatbeta_c,clique} is exact as $\tau \to 0$, yielding $\hat{\beta}_{\rm c} = 1/\alpha_{\max}^*$.
It is also exact as $\tau\to\infty$, yielding $\hat{\beta}_{\rm c} = 1/d_{\rm cl}$.

\begin{figure}
\centering
\includegraphics[width=\textwidth]{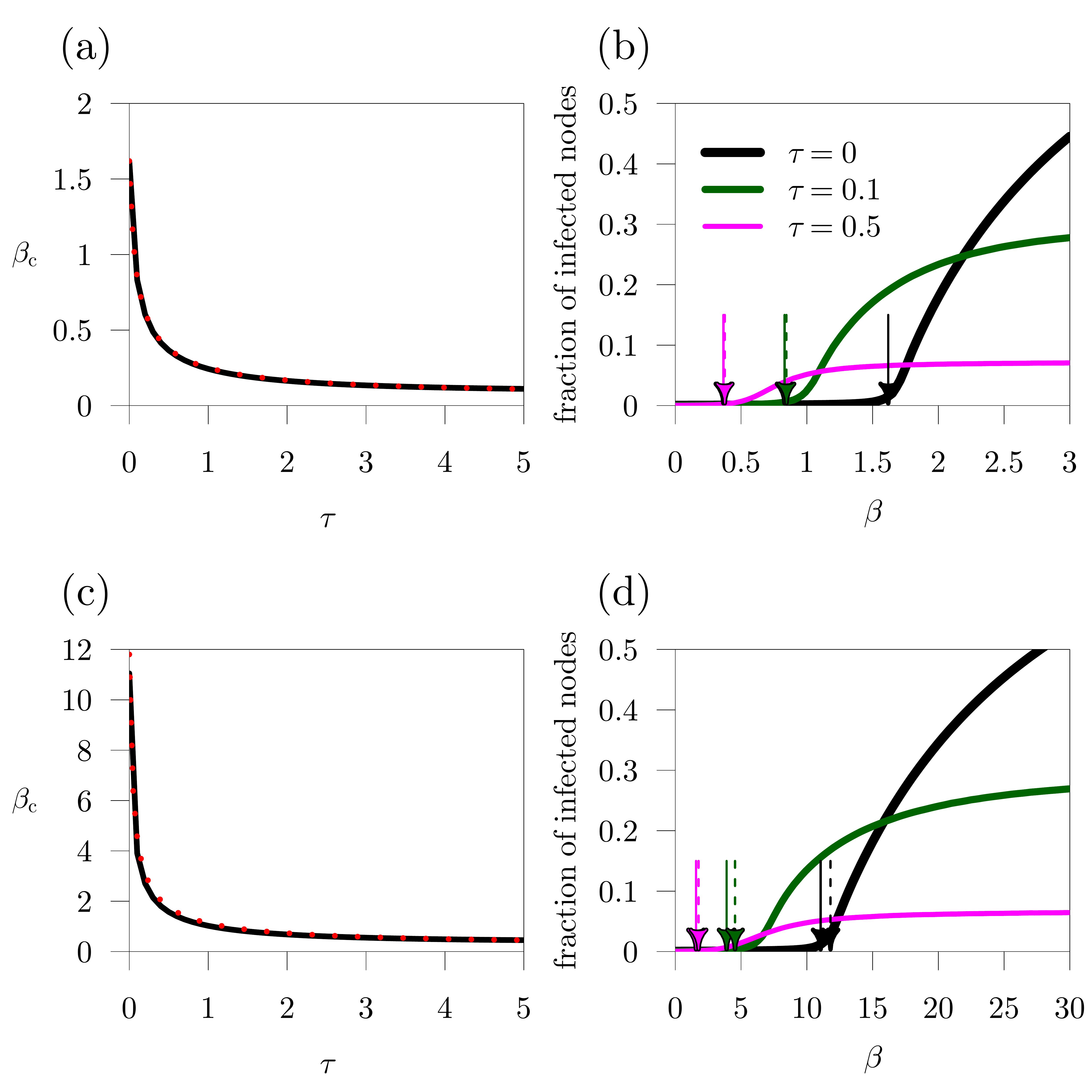}
\caption{Epidemic threshold for temporal network models. (a), (b) Temporal networks with clique snapshots. (c), (d) Activity driven model. 
(a) Epidemic threshold for the clique network model. The epidemic threshold, $\hat{\beta}_{\rm c}$, obtained from the bisection method on Eq.~\eqref{eq:effective_matrix,temporal} is shown by the solid line. The approximation, Eq.~\eqref{eq:hatbeta_c,clique}, is shown by the dotted line. The two lines severely overlap in the entire range of $\tau$.
(b) Fraction of infected nodes for the clique network model. 
The results for direct stochastic simulations are shown by the curves. The epidemic threshold predicted by Eq.~\eqref{eq:effective_matrix,temporal} and Eq.~\eqref{eq:hatbeta_c,clique} are shown by the solid and dashed arrows, respectively. 
(c) Epidemic threshold for the activity driven model. The exact value obtained from Eq.~\eqref{eq:effective_matrix,temporal} is shown by the solid line. The approximation, Eq.~\eqref{eq:hatbeta_c,simplified,star}, is shown by the dotted line.
(d) Fraction of infected nodes for the activity driven model. The epidemic threshold predicted by Eq.~\eqref{eq:effective_matrix,temporal} and Eq.~\eqref{eq:hatbeta_c,simplified,star} are shown by the solid and dashed arrows, respectively. We set $N = 2000$, $d_{\rm cl} = d_{\rm hub} = 15$, $r=1000$, and let each $a_i$ obey the power-law distribution with the probability density function $(1-\eta) a_i^{-\eta} / (1-\varepsilon^{1-\eta})$, where $\varepsilon \le a_i \le 1$. We set $\eta = 3$ and adjusted $\varepsilon$ to ensure $\langle a\rangle =0.0025$.}
\label{fig:synthetic}
\end{figure}

In Fig.~\ref{fig:synthetic}, we test the accuracy of the theory against numerical simulations using a synthetic temporal network constructed as follows. In each snapshot, every node is independently activated with probability $a_i$, which obeys a power-law distribution. Each activated node triggers a clique of size $d_{\rm cl}+1$ by involving $d_{\rm cl}$ other nodes drawn with equal probability. We allow multiedges in a snapshot. The degree of the aggregate network up to the leading order in terms of $N$ is given by $d_i^* \approx a_i d_{\rm cl} + \langle a \rangle d_{\rm cl}^2$, where $\langle a \rangle = \sum_{i=1}^N a_i/N$.
Figure~\ref{fig:synthetic}(a) suggests that Eq.~\eqref{eq:hatbeta_c,clique} (dotted line) is sufficiently close to the exact value of $\hat{\beta}_{\rm c}$ obtained through Eq.~\eqref{eq:effective_matrix,temporal} (solid line).
%
%
The small discrepancy between the exact and the approximated values are caused by the fact that cliques may overlap in the synthetic temporal networks, which the approximate theory does not assume. All these estimates accurately locate the position of the epidemic threshold obtained from direct numerical simulations (Fig.~\ref{fig:synthetic}(b)).

Different temporal networks generated by the present model can have the same aggregate network.
In Eq.~\eqref{eq:hatbeta_c,clique}, $\alpha_{\max}^*$ is the same for all temporal networks sharing an aggregate network. Therefore, the epidemic threshold depends on the temporality of networks solely through $d_{\rm cl}$. Equation~\eqref{eq:hatbeta_c,clique} implies that the epidemic threshold decreases as $d_{\rm cl}$ increases for all values of $\tau >0$. This observation implies that infection is more likely to be prevalent when snapshots are highly variable in the sense that some snapshots have many links and others have few links, as compared to when different snapshots have similar densities.

\subsection{Activity driven model}

In the temporal networks composed of cliques (section~\ref{sub:clique}), the degree of nodes is essentially homogeneous within each snapshot (i.e., $d_{\rm cl}$ or 0). In this section, we study the case in which a snapshot consists of a disjoint union of stars, allowing snapshots to be heterogeneous in the node's degree. Each star is assumed to have one hub node connected to $d_{\rm hub}$ leaves. A leaf is only adjacent to the corresponding hub. The value of $d_{\rm hub}$ is assumed to be the same for different stars and snapshots.
Different snapshots may contain different numbers of stars. As a special case of this model, we consider the discrete-time version of the activity driven model~\cite{Perra2012,Liu2014}.
In each snapshot, every node $i$ is activated with probability $a_i$ independently of other nodes. The variable $a_i$ plays a similar role to that in the case of clique snapshots but is distinct from it. For each activated node, $d_{\rm hub}$ nodes are drawn with equal probability and connected to the activated node. Although stars in a single snapshot may overlap, we consider the case in which the overlap is rare. 

As shown in Appendix~\ref{sec:epidemic threshold star}, the epidemic threshold is approximated as
\begin{equation}
\hat{\beta}_{\rm c} \approx \frac{1}{\tau \sqrt{d_{\rm hub}}} \ln \left[ 1 + \frac{\sqrt{d_{\rm hub}}}{\alpha_{\rm max}^*}  (e^\tau -1) \right].
\label{eq:hatbeta_c,simplified,star}
\end{equation}
It should be noted that both Eq.~\eqref{eq:hatbeta_c,simplified,star} and a more exact estimate given by Eq.~\eqref{eq:hatbeta_c,complex,star} (Appendix~\ref{sec:epidemic threshold star}) converge to $\beta^*_{\rm c} = 1/\alpha_{\rm max}^*$ in the limit $\tau \to 0$ and
to $1/\sqrt{d_{\rm hub}}$ in the limit $\tau \to \infty$. The high accuracy of Eq.~\eqref{eq:hatbeta_c,simplified,star} is confirmed in Fig.~\ref{fig:synthetic}(c).

Equation~\eqref{eq:hatbeta_c,simplified,star} indicates that the epidemic threshold is small for a large value of $d_{\rm hub}$. This result is consistent with that for the network model with clique snapshots. In other words, if the aggregate network is the same, temporal networks that sometimes have dense snapshots and otherwise sparse snapshots would make the epidemic threshold small, as compared to temporal networks that have similar density of links across time.

According to the heterogeneous mean field approximation~\cite{Perra2012,Liu2014}, the epidemic threshold for the activity driven model is equal to
\begin{equation}
\beta_{\rm c} \approx \frac{1}{d_{\rm hub} \left(  \langle a \rangle + \sqrt{ \langle a^2\rangle}\right)}.
\end{equation}
In fact, we obtain $\alpha_{\max}^* \approx d_{\rm hub} \left(  \langle a \rangle + \sqrt{ \langle a^2 \rangle}\right)$, yielding $\beta^*_{\rm c} = \beta_{\rm c}$ (see~Appendix~\ref{sec:activity driven aggregate} for the derivation of $\alpha_{\max}^*$). In their framework, network switching occurs sufficiently fast as compared to epidemic dynamics such that epidemic spreading is effectively occurring on the static, aggregate network. Their epidemic threshold is different from the well-known value for the configuration model \cite{Pastorsatorras2001PhysRevLett} because the aggregate network of the activity driven network is different from the configuration model having the same degree sequence. 
In contrast, the present results capture how the time scale of the network dynamics as described by the activity driven model and that of the SIS dynamics interact.

\section{Non-commuting snapshots lower the epidemic threshold\label{sec:C}}

\begin{figure}
\centering
\includegraphics[width=0.5\textwidth]{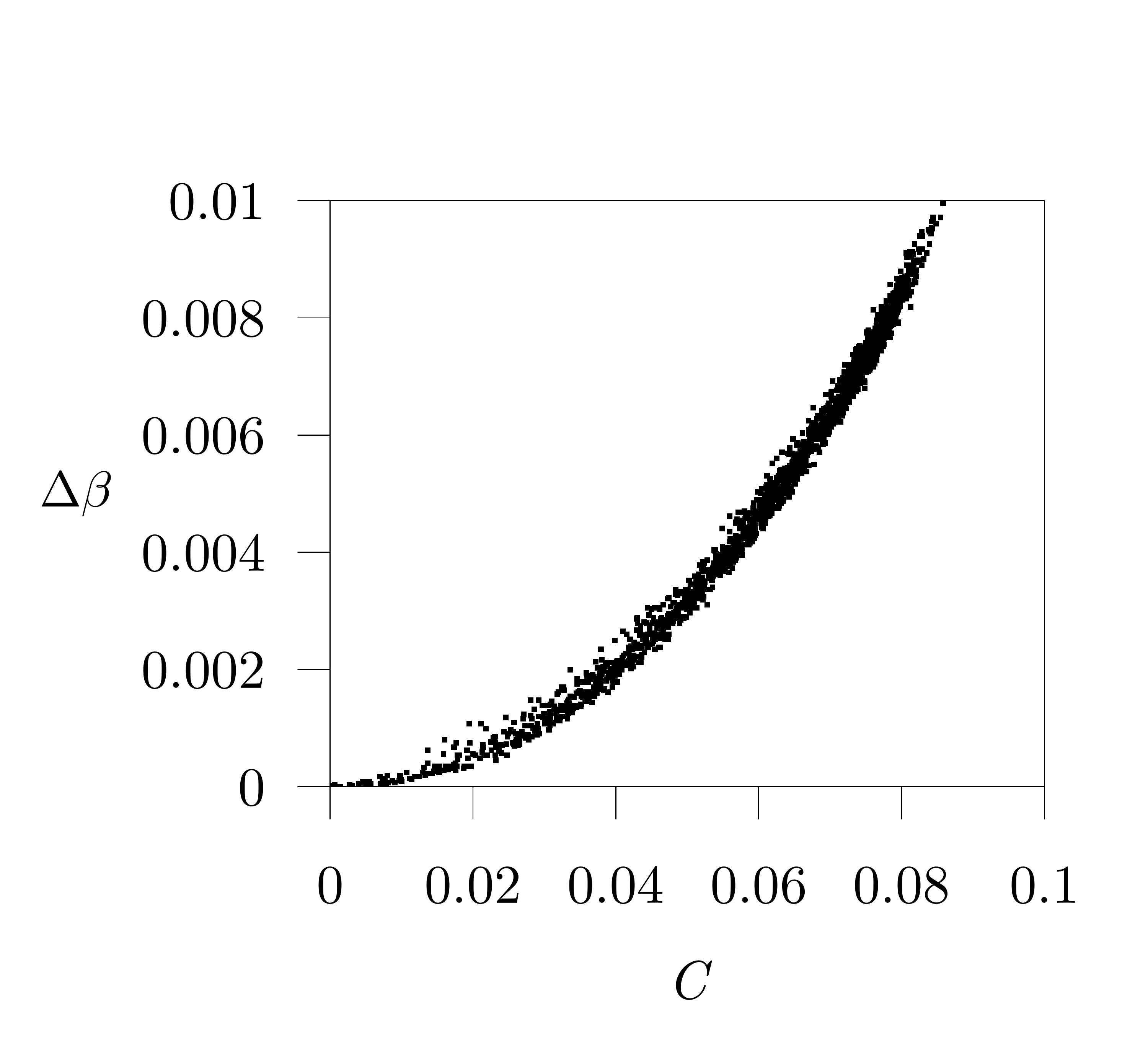}
\caption{Relationship between the epidemic threshold and the degree of non-commutativity for the synthetic temporal networks. Each point corresponds to a different sequence of snapshots manipulated from the original one such that the aggregate network is the same. The original sequence was generated with $r=50$ snapshots and $N=10$ nodes each. We used a small $N$ value because it was computationally costly to generate snapshots with larger $N$. We set $\tau = 1$.}
\label{fig:Cbeta}
\end{figure}

The amount of the shift in the epidemic threshold as we slow down the dynamics of the network (by increasing $\tau$) depends on individual temporal networks. The online message network (Fig.~\ref{fig:empirical}(a)) experiences a larger shift than the sexual contact network (Fig.~\ref{fig:empirical}(b)). In the model networks examined in section~\ref{sec:modelnetwork}, the epidemic threshold is sensitive to the change in $\tau$ when $d_{\rm cl}$ or $d_{\rm hub}$ is large (Eqs.~\eqref{eq:hatbeta_c,clique} and \eqref{eq:hatbeta_c,simplified,star}).
In this section, we propose a quantity to predict the sensitivity of the epidemic threshold, $\beta_{\rm c}$, to temporality of networks, $\tau$.

First of all, $\beta_{\rm c}$ is independent of $\tau$ if any pair of adjacency matrices of the snapshots commutes. This is because the time evolution operator for the temporal network, $T(\tau)$ (Eq.~\eqref{eq:T(S;tau),general}), and that for the aggregate network, $\exp\left[(\beta A^\ast - I)  \ell \tau \right] $, coincide in this case. To quantify the difference between $\beta_{\rm c}$ and $\beta_{\rm c}^*$ when matrices are non-commuting, we use Zassenhaus' formula~\cite{Magnus1954} given by
\begin{equation}
\exp\left[s(M_1 + M_2) \right] = \exp\left[sM_1\right] \exp\left[sM_2\right] \prod_{n \ge 2} \exp\left[s^n C_n(M_1,M_2)\right], 
\label{eq:Zassenhaus}
\end{equation}
where $M_1$ and $M_2$ are matrices, $s$ is a real number, and $\prod_{n \ge 2}$ indicates a matrix product in ascending order of the indices. Matrices $C_n(M_1,M_2)$ ($n \ge 2$) are given by linear combinations of nested commutator brackets, where the commutator bracket of $M_1$ and $M_2$ is defined by
\begin{equation}
[M_1, M_2] \equiv M_1M_2 - M_2M_1.
\end{equation}
For instance, the first two matrices are given by
\begin{equation}
C_2(M_1,M_2) = -\frac{1}{2} [M_1,M_2]
\end{equation}
and
\begin{equation}
C_3(M_1,M_2) = \frac{1}{6} \left( 2[M_2,[M_1,M_2]] + [M_1,[M_1,M_2]] \right).
\end{equation}
By iteratively applying Eq.~\eqref{eq:Zassenhaus} with $s=\tau \beta$, we obtain
\begin{align}
\exp\left[(\beta A^* - I) \ell \tau \right] &= T(\tau) \prod_{\ell^\prime=1}^{\ell - 1} \prod_{n \ge 2} \exp\left[ \left(\tau\beta\right)^n C_n\left(A^{(\ell^\prime)}, \sum_{\ell^{\prime\prime}=0}^{\ell^\prime-1} A^{(\ell^{\prime\prime})} \right) \right],
\end{align}
which relates the time evolution operator for epidemic dynamics on the aggregate network on the left-hand side to that of the temporal network, $T(\tau)$, on the right-hand side. In particular, $\mu_{\rm max}$ equals
\begin{align}
\mu_{\rm max}
& = \phi\left( \exp\left[(\beta A^\ast - I)  \ell \tau \right] \left\{ I - (\tau \beta)^2 \sum_{\ell^{\prime} = 1}^{\ell-1} C_2\left(A^{(\ell^\prime)}, \sum_{\ell^{\prime\prime} = 0}^{\ell^\prime-1} A^{(\ell^{\prime\prime})} \right) - (\tau \beta)^3 \sum_{\ell^{\prime} = 1}^{\ell-1} C_3\left(A^{(\ell^{\prime})},\sum_{\ell^{\prime\prime} = 0}^{\ell^\prime-1} A^{(\ell^{\prime\prime})} \right) +   \cdots \right\}  \right) \nonumber\\ 
& = \phi\left(  I + \left(\beta A^* - I \right) \ell \tau + \left\{ \left(\beta A^* - I\right)^2 + \frac{\beta^2}{\ell^2} \sum_{\ell^{\prime} = 1}^{\ell-1}  \sum_{\ell^{\prime\prime} = 0}^{\ell^\prime-1}  \left[A^{(\ell^\prime)},A^{(\ell^{\prime\prime})} \right]  \right\} \frac{ (\ell \tau)^2}{2} + \cdots \right).
\label{eq:mu_max,expanded}
\end{align}
Equation~\eqref{eq:mu_max,expanded} suggests that nested commutator brackets between adjacency matrices of snapshots control the difference between $\mu_{\max}$ and $\phi(\exp\left[(\beta A^\ast - I)  \ell \tau \right])$. On the basis of Eq.~\eqref{eq:mu_max_general}, this difference yields a difference between $\beta_{\rm c}$ and $\beta_{\rm c}^*$. Therefore, we define the degree of non-commutativity by
\begin{equation}
C \equiv \frac{1}{(\ell \alpha_{\max}^*)^2} \sum_{\ell^{\prime} = 1}^{\ell-1}  \sum_{\ell^{\prime\prime} = 0}^{\ell^\prime-1}  \left\Vert \left[A^{(\ell^\prime)}, A^{(\ell^{\prime\prime})} \right] \right\Vert_2,
\label{eq:C}
\end{equation}
where $\Vert \cdot \Vert_2$ is the spectral norm defined by $\Vert M \Vert_2 = \sqrt{\phi(MM^\top)}$. The multiplicative constant $1/\alpha_{\rm max}^*$ in Eq.~\eqref{eq:C} normalises the leading eigenvalue of the aggregate network to unity. If $C = 0$, all pairs of the adjacency matrices of snapshots commute, and $\beta_{\rm c}$ is independent of $\tau$. If $C >0$, at least some adjacency matrices do not commute.

We carry out numerical simulations to examine the relationship between $C$ and the epidemic threshold. First, we generate $r=50$ commuting adjacency matrices of snapshots (Appendix~\ref{sec:model C}), yielding $C=0$. Then, we manipulate $C$ by gradually swapping elements of adjacency matrices of different snapshots with the aggregate network fixed (Appendix~\ref{sec:model C}). For each sequence of adjacency matrices thus obtained, we estimate the relative change of the epidemic threshold given by $\Delta \beta_{\rm c} = (\beta_{\rm c}^\ast-\beta_{\rm c})/\beta_{\rm c}^\ast$. Figure~\ref{fig:Cbeta} indicates that $\Delta \beta_{\rm c}$ increases roughly quadratically in $C$ and suggests a high predictive power of $C$.

\begin{figure}
\centering
\includegraphics[width=\textwidth]{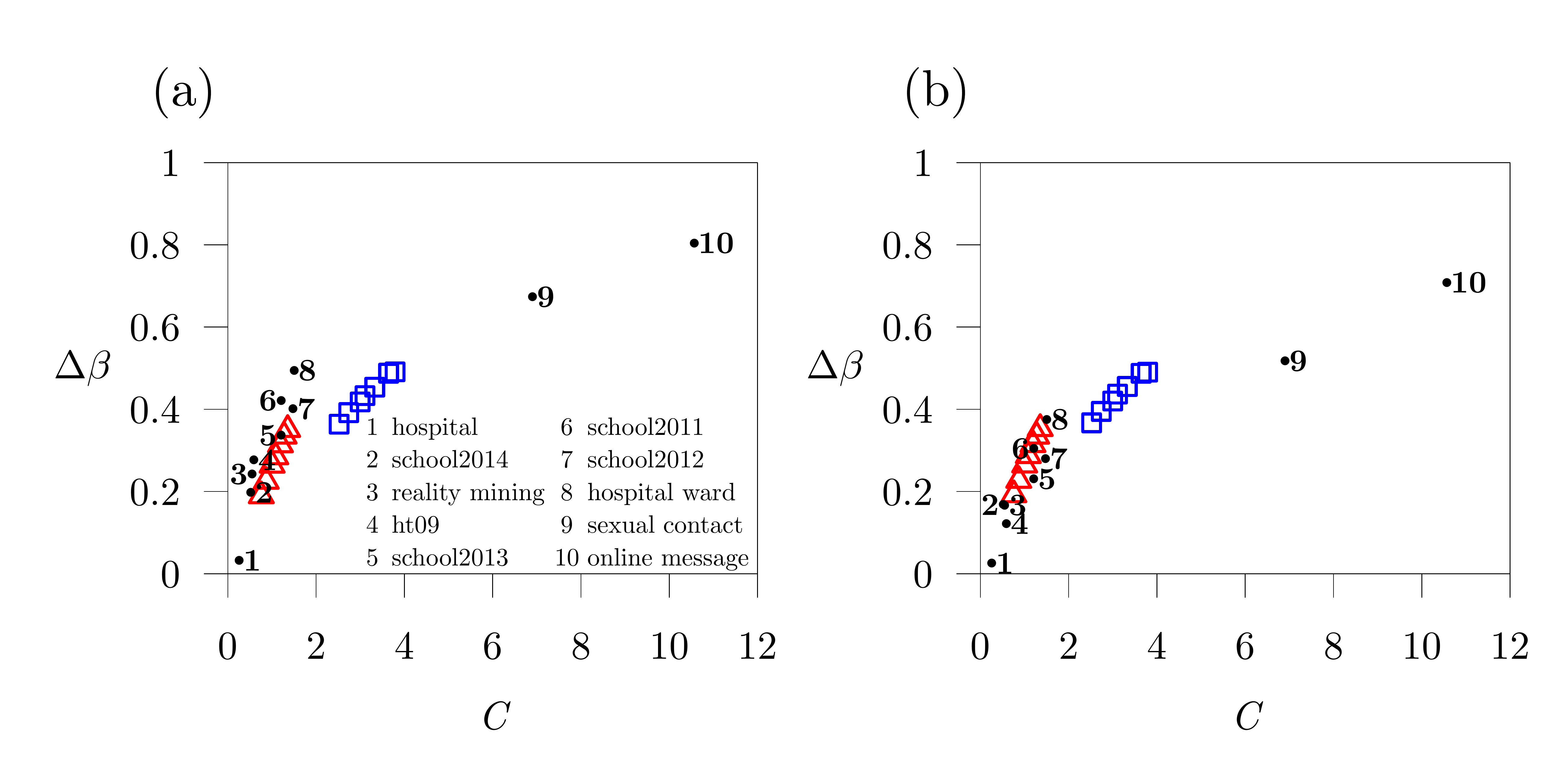}
\caption{Relationship between the shift in the epidemic threshold and the degree of non-commutativity for real temporal networks (circles), temporal networks with clique snapshots (triangles) and the activity driven model (squares). (a) Periodic boundary conditions. (b) Random sampling with replacement. Properties of the data sets are summarised in Table~\ref{tab:data}. Temporal networks with clique snapshots and the activity driven model have $N=200$ nodes, $r=200$ snapshots, and $d_{\rm cl},d_{\rm hub} \in \{4, \ldots, 10\}$. We set $\tau = 1$.
\label{fig:plot_Cdependency}
}
\end{figure}

Across several temporal network data sets summarised in Table~\ref{tab:data}, the dependency of $\Delta \beta_{\rm c}$ on $C$ is shown in Fig.~\ref{fig:plot_Cdependency}(a). The figure also contains results for temporal networks with clique snapshots and the activity driven model. These networks were generated with different values of $d_{\rm cl}$ and $d_{\rm hub}$ under the condition that the aggregate network was approximately the same within the same model (Appendix~\ref{app:varying_d}). Consistently with Fig.~\ref{fig:Cbeta}, $C$ is a strong determinant of the decrease in the epidemic threshold across various temporal networks. It should be noted that $C$ and $\Delta \beta_{\rm c}$ are strongly correlated despite different sizes of the empirical networks.

Various types of temporal correlation in empirical data are known to affect epidemic processes \cite{Karsai2011PhysRevE,Miritello2011PhysRevE,Rocha2011PlosComputBiol,Masuda2013F1000Prime}. Therefore, the epidemic threshold may be influenced by the order of snapshots, whereas $C$ is not. To examine this point, we calculated the epidemic threshold for each network using the order of snapshots given by random sampling with replacement. The results are shown in 
Fig.~\ref{fig:plot_Cdependency}(b). For these decorrelated temporal networks as well, $C$ and $\Delta \beta_{\rm c}$ are strongly correlated. Comparison between Figs.~\ref{fig:plot_Cdependency}(a) and \ref{fig:plot_Cdependency}(b) reveals that the loss of temporal correlation somewhat decreases $\Delta \beta_{\rm c}$ for all data sets.
However, $C$ is clearly a stronger determinant of $\Delta \beta_{\rm c}$ than the order of snapshots is.

\section{Discussion}

We have provided evidence that the epidemic threshold for the SIS model on temporal networks is smaller than that for the corresponding static networks for arbitrary temporal networks. We have also shown that the degree of commutativity of the adjacency matrices of the snapshot networks predicts the impact of temporality of the network on the epidemic threshold. Our results are opposite to the previous results concluding that infection in the SIS model is less likely in temporal than static networks \cite{Eames2004,Perra2012,Ferreri2014}. However, our results do not contradict theirs, for which the aggregate network obtained from the temporal network is not equal to the static network used for comparison. In contrast, we compared temporal networks with static networks such that they are the same if we ignore the temporal information in the former.

In the discrete-time SIS model on temporal networks \cite{Valdano2015PhysRevX}, the epidemic threshold is larger for the temporal network than the corresponding aggregate network when a sequence of snapshots is randomly drawn with replacement (Appendix~\ref{sec:discrete time}). This result is opposite to the current results for the continuous-time dynamics. Because a discrete-time model is a proxy to the continuous-time counterpart, which is usually more realistic, the present continuous-time framework is useful. For instance, the discrete-time SIS model implicitly assumes that, for time $\tau$, each node is allowed to make at most one transition between the infected and susceptible states. Therefore, the propagation speed is restricted, which is not the case in the continuous-time framework. In particular, the discrete and continuous-time versions only coincide when $\tau \to 0$, in which limit both reduce to the SIS model on the aggregate network.

We modelled temporal networks by switching networks. In practice, we cannot manipulate the duration of each snapshot (i.e., $\tau$) because it is specified by the data, reflecting the temporal resolution of the observation. Rather, our interpretation of $\tau$ is the relative time scale between the epidemic dynamics and network dynamics. A large $\tau$, with which the epidemic threshold decreases, corresponds to fast epidemic dynamics relative to network dynamics. The present results imply that infection is pronounced when the network varies over time (i.e., temporal network) but only slowly. In this situation, there are times when some snapshots strongly favour infection as compared to typical snapshots, and such snapshots enhance infection more than other snapshots suppress it. It should be noted that the time scale of the epidemic dynamics does not affect the equilibrium of the SIS model in the case of static networks. A future extension of the model may relax the assumption of equal duration $\tau$ of each snapshot and explore the effect of a (possibly broad) distribution of time spans. In a further step, a dependence of $\tau$ on the nodes' state may be considered to model contact avoidance and quarantine of infected nodes.

Our commutativity result opens the way to contain infection by devising the set of snapshots without changing $\tau$ or the structure of the aggregate network. In a hospital, it may be undesirable to change aggregate interactions between doctors, nurses, and patients because the amount of the interactions may be positively correlated with service quality. We can increase the epidemic threshold (therefore, less epidemic) by designing a sequence of interactions such that the corresponding adjacency matrices commute as much as possible. The method explained in Appendix~\ref{sec:model C} is useful in systematically generating commuting adjacency matrices. For a similar attempt in synchronisation dynamics using single-link snapshot networks, see~\cite{Masudapreprint}. The adjacency matrices obviously commute when the snapshots do not have any nodes in common. Therefore, designing interactions such that different sets of nodes are active at different times as much as possible may be effective at increasing the epidemic threshold. This point warrants further work. Examining the influence of $\tau$ on the epidemic threshold and the relevance of the commutativity of the adjacency matrices in more complex compartmental models of epidemic dynamics also warrants future work.


\newpage
\appendix

\section{QS method\label{sec:QS method}}

The QS method is used for computing the average prevalence in the SIS model in finite populations \cite{Oliveira2005}.
The QS method for the SIS model in static networks works as follows. We distinguish active states, which are configurations with at least one infected node, from the absorbing state, which is the configuration with no infected nodes. We keep a total of $2000$ active states in memory. After one update event in the SIS model, we are either in an active or an absorbing state. If an active state is reached, it replaces a randomly chosen state in the memory with the probability proportional to the expected time to the next update, where the proportionality constant is set to $0.5$. It should be noted that the probability does not exceed unity because the mean time to recovery is normalised to unity such that the expected time to the next update in the entire network is less than unity. If the absorbing state is reached, one active state is chosen uniformly at random from the memory to replace the absorbing state. After a transient of $10^3$ time units, the QS is calculated as the average of the system's state over the next $10^3$ time units.

We modified the QS method for temporal networks. Our implementation is slightly different from that in Ref.~\cite{Valdano2015PhysRevX}. Assume that we are currently using the $n$th snapshot. In other words, $(n-1)\tau \le t < n\tau$. The QS may depend on $t-(n-1)\tau$, i.e., the time since the beginning of the current snapshot. Therefore, for different values of $t-(n-1)\tau$ and the current snapshot, we generate a memory, i.e., a list of active states to be used when the process dies.
To this end, we divide the time window $[0,\tau)$ into those of length $\tau^{\prime}$. If $\tau < \tau^{\prime}$, we set $\tau^{\prime} = \tau$ and the time window is not divided. For every combination of the currently used snapshot and discrete time $n^\prime \tau^{\prime}$, which covers $n^{\prime} \tau^{\prime} \le t-n\tau < (n^{\prime}+1)\tau^{\prime}$ ($n^\prime = 0, 1, \ldots$), we allocate memory to store $2000$ active states. When the process is in an active state at time $t-n\tau = n^\prime \tau^{\prime}$, it replaces a randomly chosen active state in the memory corresponding to $n^\prime \tau^{\prime}$ and the current snapshot with probability 0.5. If the process reaches the absorbing state at time $t-n\tau \in [n^\prime \tau^{\prime}, (n^\prime+1) \tau^{\prime})$, the network state is replaced by an active state randomly chosen from the memory corresponding to the elapsed time $n^\prime \tau^{\prime}$ and the current snapshot. Then, the process is restarted at time $t = n\tau + n^\prime \tau^{\prime}$. After a transient of time $10^3 r\tau$, where $r$ is the number of snapshots, we calculate the steady state as the fraction of the infected nodes averaged over the $10^3 r$ subsequent measurements conducted when snapshots switch from one to another. We used $\tau^{\prime} = 0.5$ and confirmed that the results remained unaffected with $\tau^{\prime} = 0.1$.

\section{Stochastic SIS model on single-link snapshots\label{sec:1 edge}}

\begin{figure}
\begin{center}
\includegraphics[width=0.5\textwidth]{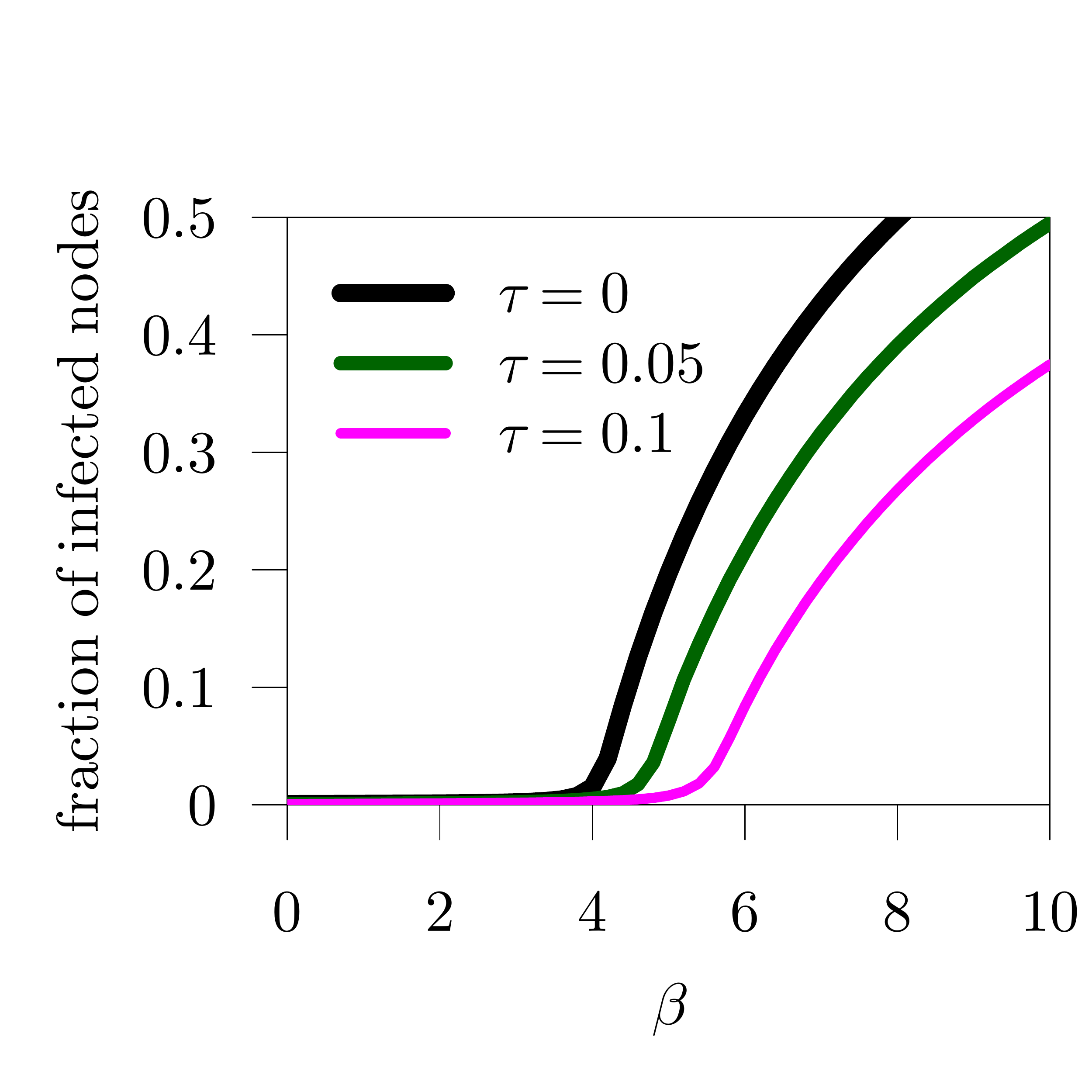}
\caption{Results of stochastic simulations on a temporal network composed of disjoint links. We set  $N=2000$ and $r=1000$. Each snapshot contains 500 disjoint links.}
\label{fig:monogamous}
\end{center}
\end{figure}

Consider a temporal network in which each snapshot contains disjoint single links drawn at random.  
The degree of every node in a snapshot is at most one. Numerical results on a temporal network generated from this model with $N=2000$ nodes and 500 links in each snapshot are shown in Fig.~\ref{fig:monogamous}. Contrary to the results shown in the main text, the epidemic threshold increases as $\tau$ increases.

This result is caused by the fact that the individual-based approximation does not work, even qualitatively, when snapshots are composed of small fragments such as disjoint single links. For such networks, stochasticity of the dynamics that the individual-based approximation does not account for plays a significant role. In short, even if the infection rate is very large, infection will die out if we apply a snapshot for long $\tau$. To understand this situation, here we analyse the stochastic SIS model on a single link $(N=2)$ by the master equation rather than by the individual-based approximation.

The SIS model with two distinguishable nodes has $2^2 =4$ states, where each node is either susceptible or infected. We assume that the two nodes are bi-directionally coupled. We denote the time-dependent probabilities of the states by $p_{\rm SS}$ (none infected), $p_{\rm IS}$ (only node
1 infected),  $p_{\rm SI}$ (only node 2 infected), and $p_{\rm II}$ (both nodes infected). It should be noted that $p_{\rm SS} + p_{\rm IS} + p_{\rm SI} + p_{\rm II}=1$. The probabilities evolve as
\begin{align}
\dot{p}_{\rm SS} &=  p_{\rm IS} + p_{\rm SI},\\
\dot{p}_{\rm IS} &= (-\beta-1) p_{\rm IS} + p_{\rm II},\\
\dot{p}_{\rm SI} &= (-\beta-1) p_{\rm SI} + p_{\rm II},\\
\dot{p}_{\rm II} &= \beta (p_{\rm IS} + p_{\rm SI}) - 2 p_{\rm II}.
\label{eq:dp_II/dt}
\end{align}
To exploit the symmetry, we consider $u=p_{\rm IS}+p_{\rm SI}$ and $r=p_{\rm IS}-p_{\rm SI}$ in place of $p_{\rm IS}$ and $p_{\rm SI}$. They evolve as
\begin{equation}
\dot{u} = (-\beta -1) u + 2 p_{\rm II}
\label{eq:du/dt}
\end{equation}
and
\begin{equation}
\dot{r} = (-\beta-1) r.
\label{eq:dr/dt}
\end{equation}

The equations for $r$, $u$, and $p_{\rm II}$ fully describe the dynamics. The solution of Eq.~\eqref{eq:dr/dt},
\begin{equation}
r(t) = r(0) e^{-(\beta+1)t},
\end{equation}
is decoupled from the dynamics of $u$ and $p_{\rm II}$. The linear dynamics composed of Eqs.~\eqref{eq:dp_II/dt} and \eqref{eq:du/dt} have the eigenvalues given by
\begin{equation}
\lambda = \frac{1}{2} (-3 - \beta \pm \kappa),
\end{equation}
with $\kappa = \sqrt{1 + 6 \beta + \beta^2}$. The corresponding left eigenvectors are given by
$\left((1-\beta\pm \kappa)/(2\beta)\; 1 \right)$.
For initial conditions $u(0)$ and $p_{\rm II}(0)$, the solution reads
\begin{align}
u(t) &= \frac{e^{(-3 -\beta) t /2}}{\kappa} \times \left\{
\left[ \beta u(0) -\frac{1}{2}(1-\beta-\kappa)p_{\rm II}(0)\right] 
   \frac{1-\beta+\kappa}{2\beta} e^{\kappa t /2} \right. \nonumber\\
& \hspace{30pt} \left. +
\left[- \beta u(0) +\frac{1}{2}(1-\beta+\kappa)p_{\rm II}(0)\right]
   \frac{1-\beta-\kappa}{2\beta} e^{-\kappa t /2}
\right\},\\
p_{\rm II}(t) &= \frac{e^{(-3-\beta) t /2}}{\kappa} \times 
 \left\{
\left[ \beta u(0) -\frac{1}{2}(1-\beta-\kappa)p_{\rm II}(0)\right] e^{\kappa t /2} \right. \nonumber \\ 
& \hspace{30pt} \left. + \left[- \beta u(0) +\frac{1}{2}(1-\beta+\kappa)p_{\rm II}(0)\right] e^{-\kappa t /2}
\right\}.
\end{align}

\begin{figure}
\begin{center}
\includegraphics[width=\textwidth]{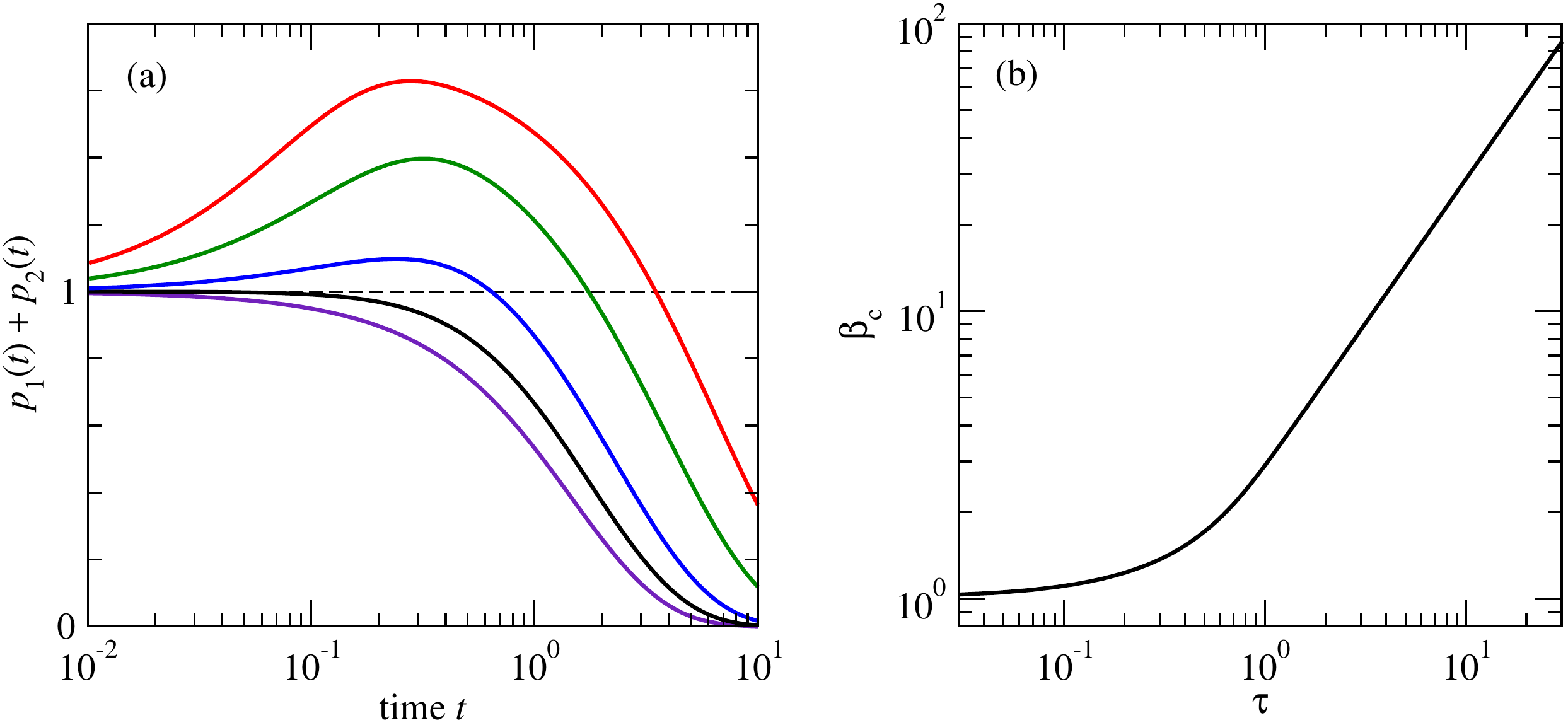}
\caption{\label{fig:1edge}
Stochastic SIS dynamics on two nodes connected by a link.
(a) Expected number of the infected nodes as a function of time when one of the two nodes is initially infected. The curves from the top to bottom correspond to $\beta=10$, 5, 2, 1, and $0.5$. (b) A lower bound for the epidemic threshold as a function of $\tau$. For each point
$(\tau,\beta_{\rm c})$, the value of $\tau$ is the time when the corresponding curve in (a) intersects with the dashed horizontal line.}
\end{center}
\end{figure}

In what follows, we assume that one node is initially infected and the other is initially susceptible, yielding $u(0) = 1$, $r(0) \in \{-1,1\}$, and $p_{\rm II}(0)=0$. Then, the solution at time $\tau$ is given by
\begin{align}
u(\tau) &= \frac{e^{(-3-\beta) \tau /2}}{2\kappa}
\left\{ 
(1-\beta+\kappa) e^{\kappa \tau /2} -
(1-\beta-\kappa) e^{-\kappa \tau /2}
\right\} \nonumber\\
&= \frac{e^{(-3-\beta) \tau /2}}{\kappa}
  \left\{(1-\beta) \sinh(\kappa \tau /2) + \kappa \cosh(\kappa \tau /2) \right\},\\
p_{\rm II} &= \frac{e^{(-3-\beta) \tau /2}}{\kappa} \left\{ 
\beta e^{\kappa \tau /2} -
\beta e^{-\kappa \tau /2}
\right\} \nonumber\\
&= \frac{2 \beta e^{(-3-\beta) \tau /2}}{\kappa} \sinh(\kappa \tau /2),
\end{align}
and
\begin{equation}
r(\tau) = r(0) e^{(-\beta-1)\tau}.
\end{equation}
We denote the probability for nodes 1 and 2 to be infected at time $\tau$ by $p_1(\tau)$ and $p_2(\tau)$, respectively. They are given by
\begin{align}
p_1(\tau) &= \frac{1}{2} r(\tau) + \frac{1}{2} u(\tau) + p_{\rm II}(\tau) \nonumber\\
&= r(0) \frac{e^{(-\beta-1)\tau}}{2} + \frac{e^{(-3-\beta) \tau /2}}{2\kappa}
\left\{ \kappa \cosh(\kappa \tau /2) +  (1+3\beta) \sinh(\kappa \tau /2) 
\right\}
\end{align}
and
\begin{align}
p_2(\tau) &= - \frac{1}{2} r(\tau) + \frac{1}{2} u(\tau) + p_{\rm II}(\tau) \nonumber\\
&= - r(0) \frac{e^{(-\beta-1)\tau}}{2} + \frac{e^{(-3-\beta) \tau /2}}{2\kappa} \left[ \kappa \cosh(\kappa \tau /2) +  (1+3\beta) \sinh(\kappa \tau /2) 
\right].
\end{align}

The expected number of infected nodes, $p_1(\tau)+p_2(\tau)$, when one node is initially infected is shown in Fig.~\ref{fig:1edge}(a). The individual-based approximation developed in the main text would predict that the number of infected nodes monotonically increases in time when $\beta$ is sufficiently large. However, Fig.~\ref{fig:1edge}(a) indicates that the infection eventually dies out even for a large $\beta$ value. Next, we estimated the epidemic threshold as the $\beta$ value at which $p_1(\tau)+p_2(\tau)=1$. This estimate gives a $\tau$-dependent lower bound 
on the epidemic threshold in temporal networks, $\beta_{\rm c}$. The epidemic can only spread if the expected number of nodes after time $\tau$ is larger than the number at time $0$. The relationship between the estimated epidemic threshold and $\tau$ is shown in Fig.~\ref{fig:1edge}(b). We find that the epidemic threshold increases as $\tau$ increases. 

\section{Relationship between the Lyapunov exponent and $\hat{\mu}_{\rm max}$\label{sec:Lyapunov proof}}

We assume random sampling with replacement of snapshots. The maximum Lyapunov exponent is given by $\lambda = \lim_{\ell \to \infty} (\ell \tau)^{-1} \ln \mu_{\rm max}$, where $\ell$ is the length of a sequence of snapshots. It should be noted that, while $\mu_{\rm max}$ is a random value for any $\ell$, the maximum Lyapunov exponent is a deterministic value owing to Theorem 2 in~\cite{Furstenberg1960}. In this section, we show that $(\ln \hat{\mu}_{\max})/\tau \ge \lambda$. 

We use Theorem 1 in Ref.~\cite{Protasov2013}. Suppose that any adjacency matrix of the snapshot is a matrix with non-negative entries and that, for each pair ($i, j$) ($1 \le i, j \le N$), there is a sequence of adjacency matrices $\{A^{(0)}, A^{(1)}, \ldots, A^{(\ell-1)}\}$ for some $\ell$ such that $T(\tau)_{ij} > 0$. Under this condition, the theorem states that for every pair ($i$, $j$) satisfying $\lim_{\ell \to \infty}  (\tau \ell)^{-1} \ln \left[ T(\tau)_{ij} \right] \neq 0$, we obtain
\begin{equation}
\lim_{\ell \to \infty} \frac{1}{\ell \tau} \ln \left[ T(\tau)_{ij} \right] = \lambda.
\label{eq:Lyapunov,entrywise}
\end{equation}
The condition of the theorem is satisfied if the aggregate network is connected.

We denote the ensemble average over the set of snapshots by $\mathbf{E}$. By using Jensen's inequality, we obtain
\begin{equation}
\mathbf{E}\left[ \lim_{\ell \to \infty} \frac{1}{\ell \tau} \ln \left[ T(\tau)_{ij} \right] \right] \le \lim_{\ell \to \infty} \frac{1}{\ell \tau} \ln \mathbf{E}\left[ T(\tau)_{ij} \right] = \lim_{\ell \to \infty} \frac{1}{\ell \tau} \ln \left[\left(\hat{T}^{\ell}\right)_{ij} \right].
\label{eq:ensemble_average,hat{T}}
\end{equation}
If $\lim_{\ell \to \infty}  (\ell \tau)^{-1} \ln \left[ T(\tau)_{ij} \right] = 0$, Eq.~\eqref{eq:ensemble_average,hat{T}} implies 
\begin{equation}
\lim_{\ell \to \infty} \frac{1}{\ell \tau} \ln \left[\left(\hat{T}^{\ell}\right)_{ij} \right] \ge 0.
\label{eq:loghatT/tau>0}
\end{equation}
If $\lim_{\ell \to \infty}  (\ell \tau)^{-1} \ln \left[ T(\tau)_{ij} \right] \neq 0$, we combine Eqs.~\eqref{eq:Lyapunov,entrywise} and \eqref{eq:ensemble_average,hat{T}} to obtain 
\begin{equation}
\lim_{\ell \to \infty} \frac{1}{\ell \tau} \ln \left[\left(\hat{T}^{\ell}\right)_{ij} \right] \ge E[\lambda] = \lambda.
\label{eq:loghatT/tau>lambda}
\end{equation}

Because the largest eigenvalue of $\hat{T}^{\ell}$ is equal to $\hat{\mu}_{\rm max}^{\ell}$ and
$\hat{T}^{\ell}$ is symmetric in the case of undirected snapshots, we obtain
\begin{equation}
\hat{\mu}_{\rm max}^{\ell} = \max_{ \vect{x}^\top \vect{x} = 1} \vect{x}^\top \hat{T}^{\ell} \vect{x}.
\label{eq:lower bound hatmu_max^ell}
\end{equation}
We distinguish two cases. 

If there is an $i$ such that $\lim_{\ell \to \infty}  (\ell \tau)^{-1} \ln \left[ T(\tau)_{ii} \right] \neq 0$, we let $\vect{x}$ have 1 in the $i$th element and zero elsewhere in Eq.~\eqref{eq:lower bound hatmu_max^ell} to obtain $\hat{\mu}_{\rm max}^{\ell} \ge (\hat{T}^{\ell})_{ii}$. By combining this inequality with Eq.~\eqref{eq:loghatT/tau>lambda}, we obtain 
\begin{equation}
\frac{1}{\tau} \ln \hat{\mu}_{\rm max} \ge \lambda.
\label{eq:hatmu_max>lambda}
\end{equation}

Otherwise, $\lim_{\ell \to \infty}  (\ell \tau)^{-1} \ln \left[ T(\tau)_{ii} \right] = 0$ for all $i$, and Eq.~\eqref{eq:loghatT/tau>0} with $i=j$ holds true. In this case, 
we let $\vect{x}$ to have $1/\sqrt{2}$ in the $i$th and $j$th elements and zero elsewhere in Eq.~\eqref{eq:lower bound hatmu_max^ell} to obtain $\hat{\mu}_{\rm max}^{\ell} \ge [(\hat{T}^{\ell})_{ii} + (\hat{T}^{\ell})_{jj} + 2(\hat{T}^{\ell})_{ij}]/2$. We find $i$ and $j$ such that $\lim_{\ell \to \infty} (\ell\tau)^{-1} \ln [(\hat{T}^{\ell})_{ij}] \ge \lambda$. By combining these equations, we obtain 
\begin{equation}
\frac{1}{\tau} \ln \hat{\mu}_{\rm max} \ge \lim_{\ell \to \infty} \frac{1}{\ell \tau} \ln \left[ \frac{(\hat{T}^{\ell})_{ii} + (\hat{T}^{\ell})_{jj} + 2(\hat{T}^{\ell})_{ij}}{2} \right] \ge \lim_{\ell \to \infty} \frac{1}{\ell\tau} \ln \left[ (\hat{T}^{\ell})_{ij} \right] \ge \lambda.
\end{equation}


\section{Epidemic threshold for temporal networks with clique snapshots\label{sec:epidemic threshold clique}}

We denote by $A_{\rm cl}$ the adjacency matrix of the network in which nodes $1, \ldots, d_{\rm cl}+1$ form
a clique and nodes $d_{\rm cl}+2, \ldots, N$ are isolated. Restricted to the clique part (i.e., principal minor of size $d_{\rm cl}+1$), the eigenvalues of $A_{\rm cl}$ are equal to $d_{\rm cl}$ and $-1$, where $-1$ has multiplicity $d_{\rm cl}$. We define a 
$(d_{\rm cl}+1) \times (d_{\rm cl}+1)$ matrix by
\begin{equation}
  \label{eq:Q,clique}
Q_{\rm cl} = \begin{pmatrix} %
  \frac{1}{\sqrt{d_{\rm cl}+1}}  & \frac{1}{\sqrt{2}}    & \frac{1}{2} \sqrt{\frac{2}{3}}  & \cdots & \frac{1}{d_{\rm cl}+1}  \sqrt{\frac{d_{\rm cl}}{d_{\rm cl}+1}} \cr
  \frac{1}{\sqrt{d_{\rm cl}+1}}  & -\frac{1}{\sqrt{2}}   & \frac{1}{2} \sqrt{\frac{2}{3}} & \cdots & \frac{1}{d_{\rm cl}+1}  \sqrt{\frac{d_{\rm cl}}{d_{\rm cl}+1}} \cr
  \frac{1}{\sqrt{d_{\rm cl}+1}}  & 0															  & -\sqrt{\frac{2}{3}}                  & \cdots & \frac{1}{d_{\rm cl}+1}  \sqrt{\frac{d_{\rm cl}}{d_{\rm cl}+1}} \cr
   \vdots                                   & \vdots                     &      \vdots                               & \ddots &  \vdots                                      \cr
  \frac{1}{\sqrt{d_{\rm cl}+1}}  & 0                             &     0                                        & \cdots &  -\sqrt{\frac{d_{\rm cl}}{d_{\rm cl}+1}}              \cr
  \end{pmatrix},
\end{equation}
where the first column of $Q_{\rm cl}$ is an eigenvector corresponding to $d_{\rm cl}$, restricted to the clique part, and the remaining columns span the eigenspace of the eigenvalue $-1$. Then, we obtain
\begin{align}
  \exp((\beta A_{\rm cl} - I) \tau)
  &= e^{-\tau} 
  \begin{pmatrix}
  	Q_{\rm cl} & 0 \cr
	0 & I
  \end{pmatrix}
  \begin{pmatrix}
    e^{\tau d_{\rm cl}\beta} &     0    \cr
            0                 &   I
  \end{pmatrix}
  \begin{pmatrix}
  	Q_{\rm cl}^\top & 0\cr
	0 & I
  \end{pmatrix} \nonumber\\ 
&= \frac{e^{-\tau}}{d_{\rm cl}+1} \left[ \left(e^{d_{\rm cl}\tau\beta}-e^{-\tau\beta}\right) A_{\rm cl} +  \frac{e^{d_{\rm cl}\tau\beta} + d_{\rm cl}e^{-\tau\beta} - d_{\rm cl}-1}{d_{\rm cl}}
\begin{pmatrix}
d_{\rm cl} I & 0\cr
0 & 0
\end{pmatrix} + (d_{\rm cl}+1)I \right].
  \label{eq:exp(tauM),calc,clique}
\end{align}
Equation~\eqref{eq:exp(tauM),calc,clique} remains true if a snapshot consists of multiple cliques.

By combining Eqs.~\eqref{eq:effective_matrix,temporal} and \eqref{eq:exp(tauM),calc,clique}, we obtain
\begin{equation}
\label{eq:hatM,clique}
  \hat{T}(\tau) = \frac{e^{-\tau}}{d_{\rm cl}+1} \left[ \left(e^{d_{\rm cl}\tau\beta}-e^{-\tau\beta}\right) A^* +  \frac{e^{d_{\rm cl}\tau\beta} + d_{\rm cl}e^{-\tau\beta} - d_{\rm cl}-1}{d_{\rm cl}} D^* + (d_{\rm cl}+1)I \right],
\end{equation}
where $A^* = \sum_A A/r$ is the aggregate network, and
$D^*$ is the diagonal matrix with degree $d^*_i$ of the aggregate network on the diagonal.

We denote the eigenvector corresponding to $\hat{\mu}_{\rm max}=1$ by $\vect{u}_{\rm cl}=(u_1, \ldots, u_N)^{\top}$. The normalisation is given by $\sum_{i=1}^N u_i = 1$. By using Eq.~\eqref{eq:hatM,clique}, we obtain
\begin{equation}
 \left(e^{d_{\rm cl}\tau\hat{\beta}_{\rm c}}-e^{-\tau\hat{\beta}_{\rm c}}\right) \langle A^*_{i \bullet},\vect{u}_{\rm cl}  \rangle +  \frac{e^{d_{\rm cl}\tau\hat{\beta}_{\rm c}} + d_{\rm cl}e^{-\tau\hat{\beta}_{\rm c}} - d_{\rm cl}-1}{d_{\rm cl}} d^*_i u_i   = (d_{\rm cl}+1)(e^\tau -1) u_i,
 \label{eq:condition,hatbeta_c,clique}
\end{equation}
where we abbreviated the $i$th row of matrix $A^*$ by $A^*_{i \bullet}$ and $\langle , \rangle$ denotes the scalar product of two vectors. Summation of Eq.~\eqref{eq:condition,hatbeta_c,clique} over $i$ yields
\begin{equation}
e^{d_{\rm cl}\tau\hat{\beta}_{\rm c}} \langle \vect{d}^*,\vect{u}_{\rm cl} \rangle +  \frac{e^{d_{\rm cl}\tau\hat{\beta}_{\rm c}} - d_{\rm cl}-1}{d_{\rm cl}} \langle \vect{d}^*,\vect{u}_{\rm cl} \rangle   = (d_{\rm cl}+1)(e^\tau -1),
 \label{eq:sum_icondition,hatbeta_c,clique}
\end{equation}
where $\vect{d}^*= (d^*_1,\ldots,d^*_N)^\top$.
By solving Eq.~\eqref{eq:sum_icondition,hatbeta_c,clique}, we obtain
\begin{equation}
\hat{\beta}_{\rm c} = \frac{1}{\tau d_{\rm cl} } \ln \left[ 1 + \frac{d_{\rm cl}}{\langle \vect{d}^*,\vect{u}_{\rm cl} \rangle } (e^\tau -1) \right].
\label{eq:beta_c,general,clique}
\end{equation}
It should be noted that $\vect{u}_{\rm cl}$ depends on $\tau$.

When $\tau = 0$, $\vect{u}_{\rm cl}$ is the eigenvector of $A^{*}$ corresponding to $\alpha_{\rm max}^*$ such that
\begin{equation}
\langle \vect{d}^*,\vect{u}_{\rm cl} \rangle= \sum_{i,j=1}^N A^*_{ij}u_i = \sum_{j=1}^N \alpha_{\rm max}^* u_j = \alpha_{\rm max}^*.
\label{eq:sum_du,clique}
\end{equation}
Therefore, $\hat{\beta}_{\rm c}$ converges to $\beta^*_{\rm c} = 1/\alpha_{\rm max}^*$ in the limit $\tau \to 0$. In the limit $\tau \to \infty$, $\hat{\beta}_{\rm c}$ converges to $1/d_{\rm cl}$, representing the fact that just one snapshot is used indefinitely long, and the epidemic threshold in a single snapshot is that of a clique.
%

Finally, we approximate
$\langle \vect{d}^*,\vect{u}_{\rm cl} \rangle \approx \alpha_{\rm max}^*$ for general $\tau$ values in Eq.~\eqref{eq:beta_c,general,clique} to obtain Eq.~\eqref{eq:hatbeta_c,clique}.
It should be noted that Eq.~\eqref{eq:beta_c,general,clique} is independent of $\langle \vect{d}^*,\vect{u}_{\rm cl} \rangle$ in the limit $\tau\to 0$ and $\tau\to\infty$ such that Eq.~\eqref{eq:hatbeta_c,clique} is exact in these two limits.


%

\section{Epidemic threshold for temporal networks with star snapshots\label{sec:epidemic threshold star}}

We denote by $A_{\rm st}$ the adjacency matrix of the network in which a star of size $d_{\rm hub}+1$ is embedded.
Without loss of generality, we assume that node 1 is the hub, nodes 2 to $d_{\rm hub}+1$ are leaves connected only to the hub, and nodes $d_{\rm hub}+2$ to $N$ are isolated. The adjacency matrix restricted to the star part (i.e., principal minor of size $d_{\rm hub}+1$) has three distinct eigenvalues $\pm \sqrt{d_{\rm hub}}$ and $0$, where the eigenvalue 0 has multiplicity $d_{\rm hub}-1$. We define a $(d_{\rm hub}+1)\times (d_{\rm hub}+1)$ matrix by
\begin{equation}
  \label{eq:Q,star}
Q_{\rm st} = \begin{pmatrix}
  \frac{1}{\sqrt{2}}  & -\frac{1}{\sqrt{2}} & 0 & 0 & \cdots & 0 \cr
  \frac{1}{\sqrt{2d_{\rm hub}}} & \frac{1}{\sqrt{2d_{\rm hub}}} &  \frac{1}{\sqrt{2}} & \frac{1}{2}  \sqrt{\frac{2}{3}} & \cdots & \frac{1}{d_{\rm hub}} \sqrt{\frac{d_{\rm hub}-2}{d_{\rm hub}-1}} \cr
  \frac{1}{\sqrt{2d_{\rm hub}}} & \frac{1}{\sqrt{2d_{\rm hub}}} & -\frac{1}{\sqrt{2}} & \frac{1}{2}  \sqrt{\frac{2}{3}} & \cdots & \frac{1}{d_{\rm hub}} \sqrt{\frac{d_{\rm hub}-2}{d_{\rm hub}-1}}\cr
  \frac{1}{\sqrt{2d_{\rm hub}}} & \frac{1}{\sqrt{2d_{\rm hub}}} &     0       &    -\sqrt{\frac{2}{3}}         & \cdots & \frac{1}{d_{\rm hub}} \sqrt{\frac{d_{\rm hub}-2}{d_{\rm hub}-1}} \cr 
    \vdots      & \vdots      &     \vdots  &                                & \ddots & \vdots \cr
  \frac{1}{\sqrt{2d_{\rm hub}}} & \frac{1}{\sqrt{2d_{\rm hub}}} &     0       &                      0         & \cdots & -\sqrt{\frac{d_{\rm hub}-2}{d_{\rm hub}-1}}\cr
  \end{pmatrix},
\end{equation}
where the first column is an eigenvector corresponding to eigenvalue $\sqrt{d_{\rm hub}}$, the second column is an eigenvector corresponding to eigenvalue $-\sqrt{d_{\rm hub}}$, and the remaining columns span the eigenspace of the zero eigenvalue. Then, we obtain
\begin{align}
  \exp\left[(\beta A_{\rm st}-I) \tau\right] 
    &= e^{-\tau} \begin{pmatrix}
Q_{\rm st} & 0\cr
0 & I
\end{pmatrix}
  \begin{pmatrix}
    e^{\tau\sqrt{d_{\rm hub}}\beta} &                        & \cr
                          & e^{-\tau\sqrt{d_{\rm hub}}\beta} & \cr
                          &                        & I
  \end{pmatrix}
 \begin{pmatrix}
  Q^\top_{\rm st} & 0\cr
  0 & I
  \end{pmatrix} \nonumber\\ 
  &= e^{-\tau}\left[ \frac{1}{\sqrt{d_{\rm hub}}} \sinh \left(\tau \sqrt{d_{\rm hub}} \beta \right) A_{\rm st} + \frac{1}{d_{\rm hub}} \left[\cosh\left(\tau \sqrt{d_{\rm hub}} \beta\right) - 1\right] A_{\rm st}^2 + I \right].
\label{eq:exp(tauM),calc,star}
\end{align}
If more than one star is embedded in a snapshot, appropriately permutated versions of Eq.~\eqref{eq:exp(tauM),calc,star} are added together. By applying Eq.~\eqref{eq:exp(tauM),calc,star} to Eq.~\eqref{eq:effective_matrix,temporal}, we obtain
\begin{equation}
\label{eq:hatM,star} 
  \hat{T}(\tau) = e^{-\tau}\left[ \frac{1}{\sqrt{d_{\rm hub}}} \sinh\left(\tau \sqrt{d_{\rm hub}} \beta\right) A^*+ \frac{1}{d_{\rm hub}} \left[\cosh\left(\tau \sqrt{d_{\rm hub}} \beta\right) - 1\right] \frac{\sum A^2}{r} + I \right],
\end{equation}
where the summation of $A^2$ runs over all possible snapshots and $r$ is the number of the possible snapshots. When $d_{\rm hub}=1$, we obtain $\sum A^2 / r = D^*$ such that Eq.~\eqref{eq:hatM,star} is consistent with Eq.~\eqref{eq:hatM,clique}.

As a special case, we consider the discrete-time version of the activity driven model~\cite{Perra2012,Liu2014}.
In each snapshot, every node $i$ is assumed to be activated with probability $a_i$ independently of the other nodes. 

If node $i$ is a hub, the probability that it connects to a node $j$ in a snapshot is equal to $d_{\rm hub}/N$. Therefore, we obtain up to the order of $1/N$
\begin{equation}
A^*_{ij} \approx \frac{\left(a_i+ a_j\right) d_{\rm hub}}{N},
\label{eq:A^*,randomstar}
\end{equation}
where we neglected the probability that both $i$ and $j$ are hubs and are connected to each other.
If node $k$ is a hub, the probability that it selects both nodes $i$ and $j$ ($\neq i$) as leaves is equal to $d_{\rm hub}(d_{\rm hub}-1)/\left[N(N-1)\right]$. Then, for $i\neq j$ we obtain
\begin{align}
\left( \frac{\sum A^2} {r} \right)_{ij} &
\approx \sum_{k=1; k \neq i,j}^N a_k  \frac{d_{\rm hub} (d_{\rm hub}-1) }{N(N-1)} \nonumber\\ 
& \approx  \langle a \rangle \frac{d_{\rm hub} (d_{\rm hub}-1)}{N},
\label{eq:A^2,randomstar}
\end{align}
where $\vect{a} = (a_1, \ldots , a_N)^\top$, $\langle a \rangle = \sum_{i = 1}^N a_i/N$, and we have neglected $O(N^{-2})$ terms. We also obtain
\begin{equation}
\left( \frac{\sum A^2}{r}\right)_{ii} = d^*_i \approx (a_i + \langle a \rangle) d_{\rm hub},
\label{eq:degree,randomstar}
\end{equation}
where $d^*_i$ is the degree of node $i$ in the aggregate network.

The epidemic threshold satisfies $\hat{\mu}_{\max} = 1$. We denote the eigenvector corresponding to $\hat{\mu}_{\rm max}=1$ by $\vect{u}_{\rm st}$. By substituting Eqs.~\eqref{eq:A^*,randomstar}, \eqref{eq:A^2,randomstar}, and \eqref{eq:degree,randomstar} in Eq.~\eqref{eq:hatM,star}, 
using $\langle d^* \rangle \equiv \sum_{i=1}^N d_i^* / N \approx 2 \langle a \rangle d_{\rm hub}$, which is derived from
Eq.~\eqref{eq:degree,randomstar}, and performing steps similar to those in Eqs.~\eqref{eq:condition,hatbeta_c,clique} and \eqref{eq:sum_icondition,hatbeta_c,clique}, we obtain
\begin{equation}
\frac{\langle \vect{d}^*, \vect{u}_{\rm st} \rangle}{\sqrt{d_{\rm hub}}} \sinh(\tau \sqrt{d_{\rm hub}} \hat{\beta_{\rm c}}) + \left( \frac{\langle d^* \rangle}{2} +\frac{\langle \vect{d}^*, \vect{u}_{\rm st} \rangle}{d_{\rm hub}}   \right) \cosh(\tau \sqrt{d_{\rm hub}}\hat{\beta}_{\rm c} ) \approx e^\tau -1 + \frac{\langle d^* \rangle}{2} +\frac{\langle \vect{d}^*, \vect{u}_{\rm st} \rangle}{d_{\rm hub}}.
\label{eq:closed_form<d*,u>,star}
\end{equation}
We substitute $\hat{\beta}_{\rm c} = (\tau \sqrt{d_{\rm hub}})^{-1} \ln \psi$ in Eq.~\eqref{eq:closed_form<d*,u>,star} to obtain
\begin{equation}
\left( \frac{\langle d^* \rangle}{2}  +\frac{\langle \vect{d}^*, \vect{u}_{\rm st} \rangle}{d_{\rm hub}} + \frac{\langle \vect{d}^*, \vect{u}_{\rm st} \rangle}{\sqrt{d_{\rm hub}}}   \right)  \psi^2 + \left( \frac{\langle d^* \rangle}{2}  +\frac{\langle \vect{d}^*, \vect{u}_{\rm st} \rangle}{d_{\rm hub}} - \frac{\langle \vect{d}^*, \vect{u}_{\rm st} \rangle}{\sqrt{d_{\rm hub}}}   \right) \approx 2\left(e^\tau -1 + \frac{\langle d^* \rangle}{2} +\frac{\langle \vect{d}^*, \vect{u}_{\rm st} \rangle}{d_{\rm hub}}\right) \psi.
\end{equation}
By solving this quadratic equation, we obtain
\begin{equation}
\hat{\beta}_{\rm c} \approx \frac{1}{\tau \sqrt{d_{\rm hub}}} \ln \frac{c_3 + \sqrt{c_3^2 - 4 c_1c_2}}{2c_1},
\label{eq:hatbeta_c,complex,star}
\end{equation}
where 
\begin{align}
c_1 =& \frac{\langle d^* \rangle}{2} +\frac{\langle \vect{d}^*, \vect{u}_{\rm st} \rangle}{d_{\rm hub}}+ \frac{\langle \vect{d}^*, \vect{u}_{\rm st} \rangle}{\sqrt{d_{\rm hub}}},\\
c_2 =& \frac{\langle d^* \rangle}{2}  + \frac{\langle \vect{d}^*, \vect{u}_{\rm st} \rangle}{d_{\rm hub}} - \frac{\langle \vect{d}^*, \vect{u}_{\rm st} \rangle}{\sqrt{d_{\rm hub}}},\\
c_3 =& 2(e^\tau -1) + \langle d^* \rangle + \frac{2 \langle \vect{d}^*, \vect{u}_{\rm st} \rangle}{d_{\rm hub}}.
\end{align}

We obtain an approximate formula for $\hat{\beta}_{\rm c}$ by replacing $\langle \vect{d}^*, \vect{u}_{\rm st} \rangle$ with $\alpha_{\rm max}^* \approx d_{\rm hub} \left(  \langle a \rangle + \sqrt{ \langle a^2\rangle}\right)$ (see Appendix~\ref{sec:activity driven aggregate} for the derivation of $\alpha_{\rm max}^*$), which is exact for $\tau=0$ and $\tau=\infty$ as in the case of temporal networks with clique snapshots (Appendix~\ref{sec:epidemic threshold clique}). We additionally simplify this formula by performing a Taylor expansion of $\hat{\beta}_{\rm c}$ in terms of $e^\tau - 1$ around $\tau = 0$. By neglecting higher order terms, we obtain Eq.~\eqref{eq:hatbeta_c,simplified,star}.

\section{Derivation of $\alpha_{\rm max}^*$ for the activity driven model\label{sec:activity driven aggregate}}

We derive the leading eigenvalue of the aggregate matrix for the activity driven model, $\alpha_{\max}^*$, where the aggregate adjacency matrix is given by Eq.~\eqref{eq:A^*,randomstar}. Each row of $A^*\vect{u}_{\rm st} = \alpha_{\max}^* \vect{u}_{\rm st}$ is given by
\begin{equation}
\alpha_{\rm max}^* u_i = \langle A^*_{i \bullet}, \vect{u}_{\rm st} \rangle \approx \frac{\left(a_i + \langle \vect{a}, \vect{u}_{\rm st} \rangle \right) d_{\rm hub}}{N}.
\label{eq:eigenvalueproblem,randomstar}
\end{equation}
Summation of Eq.~\eqref{eq:eigenvalueproblem,randomstar} over $i$ yields
\begin{equation}
\alpha_{\rm max}^* \approx \left( \langle a \rangle + \langle \vect{a}, \vect{u}_{\rm st} \rangle \right) d_{\rm hub}.
\label{eq:calc1,randomstar}
\end{equation}
By multiplying both sides of Eq.~\eqref{eq:eigenvalueproblem,randomstar} with $a_i$ and summing over $i$, we obtain
\begin{equation}
 \alpha_{\rm max}^* \langle \vect{a}, \vect{u}_{\rm st} \rangle \approx \left(\langle a^2 \rangle +  \langle a \rangle \langle \vect{a}, \vect{u}_{\rm st} \rangle\right) d_{\rm hub},
\label{eq:calc2,randomstar}
\end{equation}
where $\langle a^2 \rangle = \sum_{i=1}^N a_i^2/N$. Because entries of $\vect{u}_{\rm st}$ are all non-negative and $\vect{u}_{\rm st}$ is a non-zero vector, $\langle \vect{a}, \vect{u}_{\rm st} \rangle \neq 0$. By removing $\langle \vect{a}, \vect{u}_{\rm st} \rangle$
using Eqs.~\eqref{eq:calc1,randomstar} and \eqref{eq:calc2,randomstar}, we obtain
\begin{equation}
 \left(\alpha_{\rm max}^*\right)^2 - 2 \langle a \rangle d_{\rm hub} \alpha_{\rm max}^* +\langle a \rangle^2 d_{\rm hub}^2  - \langle a^2 \rangle d_{\rm hub}^2  \approx 0,
 \label{eq:quadratic,alpha_max,star}
\end{equation}
which results in
\begin{align}
\alpha_{\max}^* 
& \approx d_{\rm hub} \left(  \langle a \rangle + \sqrt{ \langle a^2\rangle}\right).
\label{eq:alpha_max,randomstar,appendix}
\end{align}

\section{Generating temporal networks with the same aggregate network and different values of $C$\label{sec:model C}}

To generate commuting matrices, we use the fact that symmetric matrices commute if and only if their eigenspaces coincide. We start by the matrix in which all entries are equal to unity. Its eigenvalues are $N$ and $0$. We denote the eigenvectors by $\vect{u}_i$ ($i = 1,2, \ldots$), where $\vect{u}_1 =(1\; \cdots\; 1)^{\top}/\sqrt{N}$  is the eigenvector corresponding to eigenvalue $N$. Because all entries of this matrix are positive, small changes in the eigenvalues will result in a matrix with positive entries. We choose eigenvalues 
$\lambda_1 \in [N-1,N+1]$ and $\lambda_i \in [-\varepsilon,\varepsilon]$ ($2\le i \le N$), $\varepsilon > 0$
uniformly at random and calculate a new adjacency matrix as 
\begin{equation}
\sum_{i=1}^N \lambda_i \vect{u}_i \vect{u}_i^\top.
\end{equation}
We set $\varepsilon=10$ in the numerical simulations. We repeat this procedure until we obtain $r$ matrices with positive entries. The $r$ adjacency matrices commute within themselves.

We manipulate the $r$ commuting matrices to increase the degree of commutativity, $C$ (Eq.~\eqref{eq:C}), while conserving the aggregate matrix as follows. First, we select two matrices. Then, we randomly select $(i,j)$, $1 \le i,j \le N$ and swap the $(i,j)$ entry of the two matrices. To keep both matrices symmetric, we also swap the $(j,i)$ entry of the two matrices. We repeat this procedure 2000 times during which $C$ tends to increase.
 
\section{Generation of temporal networks with different $d_{\rm cl}$ and $d_{\rm hub}$ values and the same aggregate network \label{app:varying_d}}

We generate temporal networks with clique snapshots and the activity driven model with different values of $d_{\rm cl}$ and $d_{\rm hub}$ and the common aggregate network as follows.

For temporal networks with clique snapshots with given $d_{\rm cl}$ and $\vect{a}$, the aggregate matrix up to the order of $1/N$ is given by
\begin{equation}
\left( A^{\ast} \right)_{ij}  \approx (a_i + a_j) \frac{d_{\rm cl}}{N} + \langle \vect{a} \rangle \frac{d_{\rm cl} (d_{\rm cl}-1)}{N}.
\end{equation}
Therefore, for a choice of $d_{\rm cl}^{\prime}$, we need to find $\vect{a}^{\prime}$ satisfying
\begin{equation}
d_{\rm cl}^{\prime} \left(a_i^{\prime} + \frac{1}{2} \langle \vect{a}^{\prime} \rangle (d_{\rm cl}^{\prime}-1) \right) = d_{\rm cl} \left[a_i + \frac{1}{2} \langle \vect{a} \rangle (d_{\rm cl}-1) \right].
\label{eq:clique,fixedaggregate}
\end{equation}
Equation~\eqref{eq:clique,fixedaggregate} implies
\begin{equation}
\vect{a}^{\prime} = \frac{d_{\rm cl}}{d_{\rm cl}^{\prime}} \left\{ I - \frac{1}{N} \left[\frac{d_{\rm cl}-1}{2} - \left(1+\frac{d_{\rm cl}-1}{2} \right) \frac{d_{\rm cl}^{\prime} - 1}{d_{\rm cl}^{\prime}+1}\right] J \right\} \vect{a},
\end{equation}
where $J$ is the matrix in which all entries are equal to unity.

For the activity driven model with given $d_{\rm hub}$ and $\vect{a}$, the aggregate matrix up to the order of $1/N$ is given by Eq.~\eqref{eq:A^*,randomstar}. Therefore, for a choice of $d_{\rm hub}^{\prime}$, we need to find $\vect{a}^{\prime}$ satisfying
\begin{equation}
\frac{(a_i^{\prime} + a_j^{\prime}) d_{\rm hub}^{\prime}}{N} = \frac{(a_i + a_j) d_{\rm hub}}{N},
\end{equation}
which implies
\begin{equation}
\vect{a}^{\prime} = \frac{d_{\rm hub}}{d_{\rm hub}^{\prime}} \vect{a}.
\end{equation}

In Fig.~\ref{fig:plot_Cdependency}, we chose $d_{\rm cl}$, $d_{\rm hub}=10$ and drew $a_i$ from a power-law distribution with exponent $3$ and mean $\langle a \rangle = 0.05$.  

\section{Epidemic threshold for temporal networks in discrete time\label{sec:discrete time}}

In this section, we show $\beta_{\rm c, disc} \ge \beta^*_{\rm c}$, where 
$\beta_{\rm c, disc}$ is the epidemic threshold for the discrete-time SIS model in which snapshots are randomly sampled with replacement from a given set.

For the SIS model in discrete time, the time evolution operator is given by Eq.~(8) in \cite{Valdano2015PhysRevX} as follows:
\begin{equation}
T_{\rm disc}(\tau) = \left[(1-\tau) I + \tau \beta A^{(\ell-1)} \right] \left[(1-\tau) I + \tau \beta A^{(\ell-2)} \right]  \cdots  \left[(1-\tau) I + \tau \beta A^{(0)} \right] .
\end{equation}
Because the probability of infection and recovery is given by $\tau \beta$ and $\tau$, respectively, $\tau$ must be smaller than
$\min\{ \beta^{-1}, 1\}$. The epidemic threshold, $\beta_{\rm c, disc}$, is equal to the value of $\beta$ at which the largest eigenvalue of $T_{\rm disc}(\tau)$ is equal to unity. 

We let
\begin{equation}
\hat{T}_{\rm disc} \equiv \mathbf{E}[ T_{\rm disc}(\tau)] = [(1-\tau) I + \tau \beta A^*]^{\ell}
\label{eq:hatT_disc}
\end{equation}
and denote by $\hat{\beta}_{\rm c, disc}$ the value of $\beta$ at which the largest eigenvalue of $\hat{T}_{\rm disc}$ is equal to unity. Because Eq.~\eqref{eq:hatT_disc} indicates that the largest eigenvalue of $\hat{T}_{\rm disc}$ is equal to $[(1-\tau)+\tau \beta \alpha^*_{\rm max}]^{\ell}$, we obtain
\begin{equation}
\hat{\beta}_{\rm c, disc} = \frac{1}{\alpha_{\rm max}^*} =\beta^*_{\rm c}
\label{eq:tilde_beta=beta*}
\end{equation}
regardless of $\tau$. Under the restriction that $0\le \tau\le \min\{ \beta^{-1}, 1\}$, matrices $(1-\tau) I + \tau \beta A^{(\ell^\prime)}$ are non-negative. Then, we can apply arguments similar to those in Appendix~\ref{sec:Lyapunov proof} to show that $\beta_{\rm c, disc} \ge \hat{\beta}_{\rm c, disc} (= \beta^*_{\rm c})$.

As a demonstration, we calculated the epidemic threshold for the discrete-time SIS model for the ht09 data set (Table~\ref{tab:data}). The epidemic threshold for the aggregate network was equal to $\beta^*_{\rm c} = 0.046$. With $\tau = 0.2$ and $\tau=0.5$, we obtained $\beta_{\rm c, disc} = 0.079$ and  $0.098$, respectively, confirming that the epidemic threshold increases as $\tau$ increases in the discrete-time SIS model.

\section*{Acknowledgments}
LS acknowledges the support provided through the Engineering and Physical Sciences Research Council (EPSRC) [grant number EP/G03706X/1]. LS and NM acknowledge the support provided through JST, ERATO, Kawarabayashi Large Graph Project. KK acknowledges funding from the Ram\'{o}n y Cajal program of MINECO. VME acknowledges support from MINECO (Spain) [project NOMAQ (FIS2014-60343-P)]. NM acknowledges the support provided through JST, CREST.


\end{document}